\documentclass[onecolumn,amsmath,amssymb,nofootinbib,letterpaper,11pt]{article}
\pdfoutput=1
\usepackage{jheppub}
\usepackage{ifpdf}
\usepackage[utf8]{inputenc}
\usepackage[T1]{fontenc}

\input{epsf}
\usepackage{epsfig}
\usepackage{epstopdf}
\usepackage{arcs}

\usepackage{graphicx}
\usepackage{subfigure}

\usepackage{tensor}
\usepackage{braket}
\usepackage{float}
\usepackage[dvipsnames]{xcolor}
\usepackage{amsmath}
\usepackage{morefloats}
\newcommand{\be}{\begin{equation}}
\newcommand{\ee}{\end{equation}}

\usepackage{tikz}
\usetikzlibrary{calc}   
\usetikzlibrary{topaths}
\usepackage{hyperref}
\hypersetup{
	colorlinks =WildStrawberry,
	linkcolor =red,
	pdftitle={Mimetic Inflation},
	citecolor=NavyBlue,
	filecolor=WildStrawberry,
	urlcolor=WildStrawberry,
}

\usepackage{color}

\usepackage{array}
\usepackage{bm}

\usepackage[toc]{appendix}
\usepackage{color,soul}
\usepackage{datetime}
\newcommand{\bea}{\begin{eqnarray}}
\newcommand{\eea}{\end{eqnarray}}
\newcommand{\ba}{\begin{eqnarray}}
\newcommand{\ea}{\end{eqnarray}}

\newcommand{\beq}{\begin{equation}}
\newcommand{\eeq}{\end{equation}}
\newcommand{\beqa}{\begin{eqnarray}}
\newcommand{\eeqa}{\end{eqnarray}}
\newcommand{\beqar}{\begin{eqnarray*}}
\newcommand{\eeqar}{\end{eqnarray*}}

\renewcommand{\href}[2]{#2}

\title{Multi-field Mimetic Gravity}
\author{Seyed Ali Hosseini Mansoori $^{a}$\footnote{shosseini@shahroodut.ac.ir},}
\author{Alireza Talebian $^{b}$\footnote{talebian@ipm.ir}, }
\author{Zahra Molaee $^{b}$\footnote{zmolaee@ipm.ir}, ~~~~~~~~~~~~~~}
\author{Hassan Firouzjahi $^{b}$\footnote{firouz@ipm.ir}}

\affiliation[a]{Faculty of Physics, Shahrood University of Technology, P.O. Box 3619995161 Shahrood, Iran}
\affiliation[b]{School of Astronomy, Institute for Research in Fundamental Sciences (IPM), P.O. Box 19395-5531, Tehran, Iran\vspace{0.1cm}}

\vspace{0.1cm}

\abstract{ In this paper, we extend the mimetic gravity to the multi-field setup with a 
curved field space manifold. The multi-field mimetic scenario is realized 
via the singular limit of the conformal transformation between the auxiliary and the physical metrics.   We look for the cosmological implications of the setup where it is shown that at the background level the mimetic energy density mimics the roles of dark matter. At the perturbation level, the scalar field perturbations are decomposed into the tangential and normal components with respect to the background field space trajectory. The adiabatic perturbation tangential to the background trajectory is frozen while the entropy mode  perpendicular to the background trajectory  propagates with the speed of unity. Whether or not the entropy perturbation is healthy directly depends on the signature of the field-space metric. We perform the full non-linear Hamiltonian analysis of the system with the curved field space manifold  and calculate the physical degrees of freedom verifying that the system is free from the Ostrogradsky-type ghost.  }

\preprint{
}

\begin{document}
\maketitle

\section{Introduction}

Over the years, there have been interests in theories of 
modified gravity as possible solutions to  some unsolved problems in cosmology and General Relativity (GR) like the dark matter, dark energy, the singularity problem and so on. Recently, the mimetic gravity has been proposed as a modification of GR which may mimic the roles of dark matter \cite{Chamseddine:2013kea, Chamseddine:2014vna,Chamseddine:2016uef,Chamseddine:2016ktu}. Mimetic gravity may be realized by performing a non-invertible conformal transformation of the physical metric $g_{\mu\nu}$ in the Einstein-Hilbert action  from an auxiliary metric $\tilde{g}_{\mu\nu}$ via $g_{\mu\nu}=-(\tilde{g}^{\alpha \beta} \partial_{\alpha}\phi\partial_{\beta}\phi) \tilde{g}_{\mu \nu}$
in which $\phi$ is a scalar field \cite{Deruelle:2014zza,Yuan:2015tta, Firouzjahi:2018xob}. In this way, the scalar field satisfies the constraint\footnote{Here, we use the mostly positive signature for the spacetime metric.}
\begin{equation}\label{mimetric con}
g^{\mu\nu} \partial_{\mu}\phi \partial_{\nu} \phi=-1.
\end{equation}
With this constraint, the theory reproduces the behaviour of a pressureless fluid on cosmological scale, and therefore yields a candidate of the dark matter.    
Alternatively, the mimetic gravity can be equivalently constructed by adding the above mimetic constraint via a Lagrange multiplier into the action.  For various studies on mimetic gravity setups see  Refs. \cite{Mirzagholi:2014ifa,Myrzakulov:2015kda,Arroja:2015yvd,Sebastiani:2016ras, Dutta:2017fjw, 
	Saadi:2014jfa, Gorji:2019rlm, Matsumoto:2015wja, Momeni:2015aea, Astashenok:2015qzw, Sadeghnezhad:2017hmr, Gorji:2018okn, Gorji:2019ttx, 
	Nozari:2019esz,Solomon:2019qgf, Shen:2019nyp,Ganz:2019vre, deCesare:2019pqj,Nozari:2019shm,
	deCesare:2018cts,Ganz:2018mqi,Ganz:2018vzg, Sheykhi:2019gvk, Sheykhi:2020dkm, Sheykhi:2020fqf,Astashenok:2015haa,Nojiri:2016ppu,Nojiri:2017ygt,Nojiri:2016vhu,Odintsov:2018ggm, Casalino:2018wnc, Bhattacharjee:2022lcs, Izaurieta:2020kuy, deCesare:2020swb, Kaczmarek:2021psy}.

In spite of the  fact that the original mimetic theory is free from instabilities \cite{Barvinsky:2013mea, Chaichian:2014qba}, there is no nontrivial dynamics for scalar perturbations.  This may result in caustic problem in the mimetic dark matter scenario \cite{Capela:2014xta, DeFelice:2015moy, Gumrukcuoglu:2016jbh, Babichev:2016jzg, Babichev:2017lrx}. To remedy this issue, the higher derivative term  $(\Box \phi)^2$ may be added to the action causing the scalar perturbations to propagate with a nonzero sound speed \cite{Chamseddine:2014vna,Mirzagholi:2014ifa}. Correspondingly, the mimetic setup with a general higher derivative function in the form $f(\Box \phi)$ has also been studied in Refs. \cite{Chamseddine:2016uef,Chamseddine:2016ktu}.  However, this extension of the mimetic setup with higher derivative $f(\Box \phi)$ suffers from the ghost and the gradient instabilities \cite{Ijjas:2016pad,Firouzjahi:2017txv}. To find a way of resolving the above problems, it was demonstrated in \cite{Zheng:2017qfs,Hirano:2017zox,Gorji:2017cai} that it is possible to bypass both gradient and the ghost instabilities by introducing direct couplings of the higher derivative terms to the curvature tensor of the spacetime. In this respect, more recently, the inflationary solutions in the healthy setup of extended mimetic gravity with the inclusion of higher derivative terms and the curvature tensor of spacetime  were constructed in \cite{HosseiniMansoori:2020mxj}.

The two-field extension of the mimetic gravity in the singular conformal limit of the disformal transformation has been studied in \cite{Firouzjahi:2018xob}.  By decomposing the perturbations into the adiabatic and entropy modes, it was  also shown that, similar to the original single field mimetic model, the adiabatic mode is frozen, whereas the entropy mode propagates with the sound speed equal unity with no ghost and gradient instabilities. In addition, the shift symmetry  condition imposed in two mimetic fields setup leads to the Noether current which provides a dark matter-like energy density component at the cosmological background. By imposing the shift symmetries on both scalar fields $\phi^{1}$ and $\phi^{2}$, the corresponding two-field mimetic constraint is given by
\begin{equation}
\label{eq01}
c_{1} g^{\mu\nu} \partial_{\mu}\phi^{1} \partial_{\nu} \phi^{1}+c_{2} g^{\mu\nu} \partial_{\mu}\phi^2 \partial_{\nu} \phi^2=-1 \,,
\end{equation}    
in which $c_{1}$ and $c_{2}$ are positive constant \cite{Firouzjahi:2018xob}. Without loss of generality, these constants can be absorbed into the fields through the field redefinitions $\phi^{1} \to \phi^{1}/\sqrt{c_{1}}$ and $\phi^{2} \to \phi^{2}/\sqrt{c_{2}}$, so the above mimetic constraint can be written in the covariant form,
\begin{equation}
\delta_{ab} g^{\mu\nu} \partial_{\mu}\phi^a \partial^{\mu} \phi^b =-1, \hspace{0.5cm} \text{with} \hspace{0.5cm} \phi^{a}=(\phi^{1},\phi^{2}) \,.
\end{equation}
Written in this form, one can think of $\delta_{ab}$ as a metric characterizing a flat geometry of the target space spanned by the fields $\phi^{a}=(\phi^{1},\phi^{2})$. This was a consequence of  the shift symmetry imposed in field space.

In this work we build upon \cite{Firouzjahi:2018xob} and consider a setup of multi-field mimetic gravity with a curved field space metric $G_{ab}(\phi^c)$. As such, the shift symmetry assumption is violated.  The dynamics of such multi-field models with curved field-space manifold have been extensively studied in recent years mainly in the context of inflation, dark energy, primordial non-Gaussianity and related areas \cite{Gong:2011uw,Gong:2016qmq,Gong:2016qpq,Cespedes:2012hu,Achucarro:2016fby, Mizuno:2017idt, Fumagalli:2019noh, Garcia-Saenz:2019njm,Achucarro:2019mea,Akrami:2020zfz}. 

The rest of the paper is organized as follows. In Section \ref{model} we present the multi-field extension of the original mimetic scenario.  In Section \ref{sec3} we study the cosmological implications of the setup both at the background and the perturbations levels. 
In Section \ref{AH}, we perform the Hamiltonian analysis at the full non-linear level using the Arnowitt-Deser-Misner (ADM) decomposition \cite{arnowitt2008republication}. The conclusions are presented in Section \ref{sec5} while some technicalities of the analysis are relegated to the appendices.

\section{The Model}
\label{model}

In this section, we build upon the analysis of \cite{Firouzjahi:2018xob} and
construct the mimetic setup for the general case of multi-field with a 
curved field space metric $G_{a b}$. We comment that the setup of  \cite{Firouzjahi:2018xob}  was dealing with a two-field setup with a flat field space, i.e. $G_{a b} = \delta_{a b}$,  with the constraint given by Eq. (\ref{eq01}). 

To do this, let us first assume a general conformal transformation between the physical metric $g_{\mu \nu}$ and the auxiliary metric $\tilde{g}_{\mu \nu}$ as
\begin{equation}\label{conft}
g_{\mu \nu}=A(\phi^{a},X) \tilde{g}_{\mu \nu}  \,,
\end{equation}  
where $X=G_{ab} \tilde{g}^{\mu \nu}\partial_{\mu}\phi^{a}\partial_{\nu}\phi^{b}$ in which $G_{ab}(\phi^{c})$ is the metric of the field-space manifold\footnote{The natural generalization of disformal transformation in the case of multi-field mimetic gravity can be expressed by
\begin{equation}
g_{\mu \nu}=A(\phi^{a},X) \tilde{g}_{\mu \nu}+C(\phi^{a},X)G_{ab}\partial_{\mu}\phi^{a}\partial_{\nu}\phi^{b} \,.
\end{equation}
Note that we take $C=0$ in the conformal case. }. Demanding that the determinant of $g_{\mu \nu}$  to be non-zero, its inverse metric is given by $g^{\mu \nu}=A^{-1} \tilde{g}^{\mu \nu}$.
In order to find if the above transformation is invertible, we need to look at the eigenvalue equation for the determinant of the Jacobian \cite{Zumalacarregui:2013pma}, i.e.,  
\begin{equation}
\Big(\frac{\partial g_{\mu \nu}}{\partial \tilde{g}_{\mu \nu}}-\kappa^{(n)} \delta_{\mu}^{\alpha} \delta_{\nu}^{\beta}\Big)\xi_{\alpha \beta}^{(n)}=0 \,,
\end{equation}
in which $\kappa^{(n)}$ and $\xi^{(n)}_{\mu \nu}$ are the eigenvalues and the associated eigentensors, respectively. Therefore, the eigenvalues equation for the conformal transformation (\ref{conft}) can be obtained to be
\begin{equation}
\Big(A-\kappa^{(n)}\Big)\xi_{\mu \nu}^{(n)}-A_{,X}\left\langle \xi^{(n)} \right\rangle_{X} \tilde{g}_{\mu \nu}=0   \,,
\end{equation}
where we have defined $\left\langle \xi^{(n)} \right\rangle_{X}\equiv G_{ab} \partial^{\alpha} \phi^{a} \partial^{\beta} \phi^{b} \xi_{\alpha \beta}^{(n)}$. The above relation can be proved by using the following identities;
\begin{equation}
\frac{\partial g_{\mu \nu}}{\partial \tilde{g}_{\alpha \beta}}=A \delta_{\mu}^{\alpha} \delta_{\nu}^{\beta}+ \tilde{g}_{\mu \nu} \frac{\partial A}{\partial \tilde{g}_{\alpha \beta}} \,,
\end{equation}
where
\begin{equation}
\frac{\partial A}{\partial \tilde{g}_{\alpha \beta}}=G_{ab}\delta^{\alpha}_{\rho} \delta_{\sigma}^{\beta} \partial^{\rho} \phi^{a} \partial^{\sigma}\phi^{b} A_{,X} \,.
\end{equation}
There are usually two kind of solutions for this eigenvalues which are called as the \textit{conformal type solution} and the \textit{kinetic type solution}. In the case of conformal type solution, the eigenvalues and eigentensor are given by
\begin{equation}\label{eigenv2}
\kappa^{(C)}=A \hspace{0.5cm} \text{with} \hspace{0.5cm} A_{,X} \left\langle \xi^{(C)} \right\rangle_{X}=0 \,.
\end{equation} 
Clearly, this kind of eigenvalue solution is degenerate with multiplicity of 9 since the eigentensors are limited by the above (single) constraint. On the other hand, for the kinetic type eigenvalues solution, we have
\begin{equation}\label{eigenv1}
\kappa^{(K)}=A-X A_{,X}  \,,
\end{equation}
with the eigentensor  being proportional to the metric tensor, i.e. $\xi^{(K)}_{\mu \nu}=\tilde{g}_{\mu \nu}$. 

Now we are interested in finding the singular limit of the conformal transformation (\ref{conft}) by demanding that the eigenvalues (\ref{eigenv1}) and/or (\ref{eigenv2}) to be zero. In the case of conformal type solution, the singular limit gives us $A=0$ which is not allowed. However,  for the kinetic type solution, the following constraint will be imposed on $A$ in the singular limit,
\begin{equation}
A=X A_{,X} \,.
\end{equation}
The nontrivial solution of the above differential equation for the conformal factor $A$ is
\begin{equation}
\label{AOmega}
A=- {\Omega^{-1}} X \,,
\end{equation}
in which $\Omega$ is a constant.  Therefore, $A$ is obtained to be a linear function of $X$. There are two options for the sign of $\Omega$. If we assume that the mimetic fields are timelike, which is the case for the cosmological background, then $X<0$ so we need $\Omega>0$. On the other hand, if the mimetic fields are spacelike, as for example in black hole background studied in \cite{Gorji:2020ten}, then $\Omega <0$. We are interested in cosmological implications of the multi-field mimetic gravity so we assume $\Omega>0$.

It is worth mentioning that one can not write down $\tilde{g}_{\mu \nu}$ as a function of $g_{\mu \nu}$ due to the nature of the above singular limit. By contracting both sides of the inverse metric relation, $g^{\mu \nu}=-{\Omega}\, X^{-1} \tilde{g}^{\mu \nu}$ with $G_{ab} \partial^{\mu} \phi^{a} \partial_{\nu} \phi^{b} $, one obtains
\begin{equation}\label{mimeticcon}
g^{\mu \nu} G_{ab} \partial_{\mu} \phi^{a} \partial_{\nu} \phi^{b} =-\frac{{\Omega} G_{ab} \partial_{\mu} \phi^{a} \partial_{\nu} \phi^{b} \tilde{g}^{\mu \nu}}{ X} \,,
\end{equation}
which yields
\begin{equation}\label{mimeticcons1}
g^{\mu \nu} G_{ab} \partial_{\mu} \phi^{a} \partial_{\nu} \phi^{b}= G_{ab} \partial_{\mu} \phi^{a} \partial^{\mu} \phi^{b}=-\Omega \,.
\end{equation}
This is nothing but  the mimetic constraint extended to curved multi-field  manifold. 

The above constraint can be applied to the theory via a Lagrange multiplier. For the single field mimetic gravity setup, it has been shown that the conformal transformation of the metric to an auxiliary metric is equivalent to adding a Lagrange multiplier to the action \cite{Golovnev:2013jxa}. But for the multi-field case  in curved space, this conclusion might be unclear. In the next subsection, we demonstrate that these two approaches are equivalent in the multi-field case as well.


\subsection{Equivalent Action with a Lagrange Multiplier}
\label{equivalent}

Here, our aim is to show that the action constructed from the physical metric \eqref{conft} is equivalent with the action in which the mimetic constraint \eqref{mimeticcons1} is added to it through a Lagrange multiplier in the following form
\begin{equation}\label{action1}
S=\int d^{4} x \sqrt{g} \Big[\frac{M_P^2}{2}R+\lambda \big(G_{ab}\partial_{\mu} \phi^{a} \partial^{\mu} \phi^{b}+\Omega \big)-V(\phi^{a}) \Big] \, ,
\end{equation}
in which $M_P$ is the reduced Planck mass, $\lambda$ is the Lagrange multiplier and
$V(\phi^a)$ is the potential added for the later cosmological purposes.  
As mentioned, the equivalency of the two actions for the case of single field mimetic scenario was shown in \cite{Golovnev:2013jxa} and we follow its logic here (with some new technicalities coming from  multi-field effects). 

From Eqs.~\eqref{conft} and \eqref{AOmega}, we consider the physical metric $g_{\mu\nu}$ as a function of auxiliary $g_{\mu\nu}$ and scalar fields $\phi^a$ as
\begin{align}
\label{gtildeg}
g_{\mu\nu} = -\Omega^{-1} \left(
G_{ab} \tilde{g}^{\alpha \beta}\partial_{\alpha}\phi^{a}\partial_{\beta}\phi^{b}
\right)\tilde{g}_{\mu\nu} \equiv -\Omega^{-1} \, X \, \tilde{g}_{\mu\nu} 
\end{align}
Clearly, the metric $g_{\mu\nu}$ is invariant under the conformal transformation of the auxiliary metric $\tilde{g}_{\mu\nu}$. 

The action constructed from the physical metric $g_{\mu\nu}$ can be considered as a function of scalar fields $\phi^a$ and the auxiliary metric $\tilde{g}_{\mu\nu}$ as follows
\begin{align}
\label{action-1}
S=\int d^4x \sqrt{-g(\tilde{g}_{\mu\nu},\phi^a)} \bigg( \dfrac{M_P^2}{2}
R \big(g_{\mu\nu}(\tilde{g}_{\mu\nu},\phi^a) \big)+{\cal L}_m
\bigg)
\end{align}
where ${\cal L}_m$ represents the Lagrangian density of the matter sector, which for 
our case is just the potential $V(\phi^a)$. 

By taking the variation of the action with respect to the physical metric $g_{\mu\nu}$, we arrive at
\begin{align}
\label{var1}
\delta S = \dfrac{1}{2} \int d^4x \sqrt{-g} \, \bigg( M_P^2
{\cal G}^{\mu\nu}-T^{\mu\nu}
\bigg) \delta g_{\mu\nu}  \, ,
\end{align}
in which ${\cal G}^{\mu\nu}$ is the Einstein tensor and $T^{\mu\nu}$ is the energy-momentum tensor associated to ${\cal L}_m$. 

From Eq.~\eqref{gtildeg} the variation $\delta g_{\mu\nu}$ can be written in terms of the variation of the auxiliary metric $\delta \tilde{g}_{\mu\nu}$ and the variation of scalar field $\delta \phi^a$  as follows
\begin{align}
&\delta g_{\mu \nu} = -\dfrac{X}{\Omega} \delta \tilde{g}_{\mu\nu} -\dfrac{\delta X}{\Omega} \tilde{g}_{\mu\nu}
\\
&= -\dfrac{X}{\Omega} \, \delta \tilde{g}_{\mu\nu} - \dfrac{\tilde{g}_{\mu\nu}}{\Omega} \bigg(
X G_{ab,c} \, 
\delta \phi^c + G_{ab}\Big[
-\tilde{g}^{\kappa \alpha}\tilde{g}^{\rho \beta} \delta \tilde{g}_{\alpha \beta} \partial_\kappa \phi^a \partial_\rho \phi^b + 2\tilde{g}^{\kappa \rho} \partial_\kappa \delta \phi^a \partial_\rho \phi^b
\Big]
\bigg)
\nonumber\\
&=-\dfrac{X}{\Omega} \, \delta \tilde{g}_{\alpha \beta}\,\bigg( \delta_{\mu}^{\alpha}\delta_{\nu}^{\beta} + \dfrac{G_{ab}}{\Omega} g_{\mu\nu} g^{\kappa \alpha}g^{\rho \beta}  \partial_\kappa \phi^a \partial_\rho \phi^b \bigg) - \dfrac{g_{\mu\nu}}{\Omega} \bigg(
G_{ab,c} \, X^{ab} \delta \phi^c + 2G_{ab}g^{\kappa \rho} \partial_\kappa \delta \phi^a \partial_\rho \phi^b \bigg)
\nonumber
\end{align}
where $X^{ab} \equiv  g^{\alpha\beta} \partial_\alpha \phi^a \, \partial_\beta\phi^b$. 

Thus  the corresponding equations of motion from the variation in Eq. (\ref{var1})
are obtained to be
\begin{align}
\label{GT}
{\cal G}_{\mu\nu}-T_{\mu\nu} + \Omega^{-1} \,({\cal G}-T) \, G_{ab}  \partial_\mu \phi^a \partial_\nu \phi^b = 0 \,,
\\
\label{lambda1}
({\cal G}-T)G_{ab,c} \, X^{ab}  - 2 \nabla _\kappa \bigg(({\cal G}-T)G_{ac} \nabla^\kappa \phi^a \bigg) =0 \,.
\end{align}
By taking the trace of Eq.\eqref{GT}, we have
\begin{align}
({\cal G}-T) \bigg(
G_{ab}  \partial_\mu \phi^a \partial^\mu \phi^b +\Omega
\bigg) = 0 \, .
\end{align}
When ${\cal G}-T \neq 0$, this equation yields the mimetic constraint \eqref{mimeticcons1}.

On the other hand, instead of working with the action (\ref{action-1}) with the physical metric 
$g_{\mu \nu}$ treated as a function of the auxiliary metric $\tilde{g}_{\mu \nu}$ and the scalar fields $\phi^a$, we can equivalently implement the relation \eqref{gtildeg} 
to the total action with a set of Lagrange multipliers and treat $g_{\mu \nu}$ as an independent field, i.e. 
\begin{align}
\label{action}
S = \int {\rm d}^4x \sqrt{-g} \bigg[ \dfrac{M_P^2}{2}
R+\Lambda^{\mu\nu}\Big(
g_{\mu\nu}+\Omega^{-1} X \tilde{g}_{\mu\nu}
\Big) + {\cal L}_m
\bigg] \,.
\end{align}
Here $\Lambda^{\mu\nu}$ is a set of Lagrangian multipliers added to incorporate 
the condition (\ref{action-1}) for all components of metric field. 
Consequently,  the variation of action with respect to $\Lambda_{\mu\nu}$ yields the constraint \eqref{gtildeg}. Note that despite the apparent similarity, the constrained action (\ref{action}) is different than the constrained action (\ref{action1}). More specifically, the single Lagrange multiplier $\lambda$ in action  (\ref{action1}) enforces the single constraint Eq. (\ref{mimeticcons1})
while the Lagrange multipliers $\Lambda^{\mu\nu}$ implement the relation \eqref{gtildeg} for all components of the physical metric. 

Now the variation of  action (\ref{action}) is given by
\begin{align}
\label{var2}
\delta S &=  \dfrac{1}{2} \int {\rm d}^4x \sqrt{-g} \bigg[ \big(
{\cal G}^{\mu\nu}-T^{\mu\nu}+\Lambda^{\mu\nu}
\big)\delta g_{\mu\nu} + \Omega^{-1} \Lambda^{\mu\nu} \delta \tilde{g}_{\mu\nu}  X+ \Omega^{-1} \Lambda^{\mu\nu} \tilde{g}_{\mu\nu} \delta X \bigg]
\nonumber\\
&=\dfrac{1}{2} \int {\rm d}^4x \sqrt{-g} \bigg[ \big(
{\cal G}^{\mu\nu}-T^{\mu\nu}+\Lambda^{\mu\nu}
\big)\delta g_{\mu\nu} 
+\Omega^{-1}  \bigg(
\Lambda \, G_{ab,c} X^{ab}-2\nabla_\kappa (\Lambda G_{ac} \nabla^{\kappa}\phi^a)
\bigg)\delta \phi^c
\bigg]
\nonumber\\
&\hspace{2.8cm}+ \Omega^{-1} \Lambda^{\mu\nu}X \bigg(
\delta_{\mu}^\alpha\delta_{\nu}^\beta+\Omega^{-1} G_{ab} g_{\mu\nu} g^{\kappa \alpha}g^{\rho \beta}  \partial_\kappa \phi^a \partial_\rho \phi^b
\bigg)\delta \tilde{g}_{\alpha\beta}  \, ,
\end{align}
where $\Lambda \equiv\Lambda^{\mu\nu}g_{\mu\nu}$ and we have set the boundary contribution to zero by demanding that the variation of $\delta \phi^b$ vanishes at infinity. In addition, use was made from the following relation,
\begin{align}
\delta X &= G_{ab,c} \, \tilde{g}^{\alpha\beta} \partial_\alpha \phi^a \, \partial_\beta\phi^b \delta \phi^c + G_{ab}\Big(
-\tilde{g}^{\kappa \alpha}\tilde{g}^{\rho \beta} \delta \tilde{g}_{\alpha \beta} \partial_\kappa \phi^a \partial_\rho \phi^b + 2\tilde{g}^{\kappa \rho} \partial_\kappa \delta \phi^a \partial_\rho \phi^b
\Big)
\\
&= - \Omega^{-2} X^2 \, G_{ab}  g^{\kappa \alpha}g^{\rho \beta}  \partial_\kappa \phi^a \partial_\rho \phi^b \delta \tilde{g}_{\alpha \beta} - \Omega^{-1} X \bigg(
G_{ab,c} \, X^{ab} \delta \phi^c + 2G_{ab}g^{\kappa \rho} \partial_\kappa \delta \phi^a \partial_\rho \phi^b \bigg) \, . \nonumber
\end{align}
Now the variations in Eq. (\ref{var2}) with respect to the physical metric $g_{\mu\nu}$, the auxiliary metric $\tilde{g}_{\mu\nu}$ and the scalar fields $\phi^a$, respectively lead to 
\begin{align}
\label{GTL}
{\cal G}_{\mu\nu}-T_{\mu\nu} +\Lambda_{\mu\nu}=0 \,,
\\
\label{L}
\Lambda_{\mu\nu}+\Omega^{-1} \Lambda G_{ab}    \partial_\mu \phi^a \partial_\nu \phi^b=0 \,,
\\
\label{lambda}
\Lambda G_{ab,c} \, X^{ab}  - 2 \nabla _\kappa \bigg(\Lambda G_{ac} \nabla^\kappa \phi^a \bigg) =0 \,.
\end{align}
Interestingly, the trace of Eq. \eqref{L} leads to the mimetic constraint \eqref{mimeticcons1}. Moreover, by taking the trace from Eq. \eqref{GTL}, we find that $\Lambda=T-{\cal G}$. Therefore, using this finding and substituting Eq.\eqref{L} into \eqref{GTL}, the equation of motion \eqref{GT} can be reproduced. In addition, it's straightforward to confirm that Eq. \eqref{lambda} coincides with \eqref{lambda1}. 

Now, solving for $\Lambda_{\mu\nu}$ from Eq.~\eqref{L} and plugging it into the action \eqref{action} yields
\begin{align}
\label{action2}
S &= \int {\rm d}^4x \sqrt{-g} \bigg[ \dfrac{M_P^2}{2}
R+\lambda \bigg( G_{ab} \partial^\mu \phi^a \partial_\mu \phi^b+\Omega
\bigg) +{\cal L}_m
\bigg] \, ,
\end{align}
which is the same as action \eqref{action1} with ${\cal L}=-V(\phi^a)$ and 
$\lambda \equiv \Omega^{-1} \Lambda$. 

It is worth mentioning that the action \eqref{action1} reduces to that of \cite{Firouzjahi:2018xob} when $G_{ab}$ is constant. 
Now, by rescaling the fields via $\phi^a \rightarrow \sqrt{\Omega}\,  \phi^a$, we can absorb the effects of the constant $\Omega$ into the Lagrange multiplier. Therefore, from now on we set $\Omega=1$ without loss of generality.

Note that in Ref. \cite{Firouzjahi:2018xob}  the shift symmetry for two scalar fields in the absence of the potential function was imposed so 
the metric $G_{ab}$ was  diagonal with constant elements. However, here we work in a curved field space manifold with the metric $G_{ab}(\phi^c)$ so the shift symmetry is explicitly broken.  Furthermore, in the absence of shift symmetry, we have allowed for the potential term $V(\phi^a)$ as well.

\section{Multi-Field Mimetic Cosmology }\label{sec3}

The goal of this section is to write down the background equations and the quadratic action of cosmological perturbations in  a covariance form. The value of the scalar fields $\phi^{a}(x^{\mu})$ at a given location in spacetime consists of the homogeneous background value, $\phi^{a}_{0}$, and the gauge dependent fluctuations, $\delta \phi^{a}$. The fluctuations $\delta \phi^{a}$ describe a finite coordinate displacement from the classical trajectory so they are not covariant. This motivates the construction of a vector field $Q^{a}$ in order to write down the field fluctuations in a covariant form.  

The two points like $\phi_0^a(t)$ and $\phi^a = \phi_0^a + \delta\phi^a$ are connected by a unique geodesic with respect to the field space metric $G_{ab}$  \cite{Gong:2011uw,Elliston:2012ab}.
This geodesic is parametrized by $\varepsilon$, such that $\phi^{a}(\varepsilon=0)=\phi^{a}_{0}$ and $\phi^{a}(\varepsilon=1)=\phi^{a}_{0}+\delta \phi$. 
These boundary conditions determine a unique vector $Q^{a}$ which connects the two scalar field values in such a way that $\left. D_\varepsilon\phi^a \right|_{\varepsilon=0} = Q^a \ $ where $D$ is 
 the covariant derivative with respect to the field space metric $G_{ab}$. 
 Therefore, one can expand $\delta \phi^a$ as a power series of $Q^{a}$ as \cite{Gong:2011uw,Elliston:2012ab},
\begin{equation}
\label{eq:mapping2}
 \delta\phi^a = Q^a - \frac{1}{2}\Gamma^a_{bc}Q^bQ^c + \frac{1}{6} \left( \Gamma^a_{de}\Gamma^e_{bc} - \Gamma^a_{bc;d} \right) Q^bQ^cQ^d + \cdots \, .
\end{equation}
in which $\Gamma^a_{bc}$ represents the Christoffel symbol associated with the metric $G_{ab}$. 

Note that at linear order, the field fluctuations $\delta\phi^a$ and the vector $Q^a$ are identical but at higher orders they are different. 
Thus, in the covariant manner, we must write the equations in terms of $Q^a$. 
In addition to scalar fields's perturbations, we need to perturb the metric components and the
Lagrange multiplier. 

The cosmological background in the absence of perturbations is given by the FLRW metric 
\ba
ds^2 = -d t^2 + a(t)^2 d {\bf x}^2
\ea
in which $a(t)$ is the scale factor.  Now denoting the components of the full metric via $g_{00}= -{\mathcal N}^2 + \beta_i \beta^i, g_{0i}= \beta_i, g_{ij} = \gamma_{ij}$,  the scalar perturbation parts of the metric at linear order are defined as
\begin{align}\label{pertubation}
\nonumber \mathcal N & = 1 + \alpha \, ,
\\
\beta_i & = B_{,i}  \, ,
\\
\nonumber \gamma_{ij}& =a^{2} e^{2 \psi} \delta_{ij}
\end{align}
where in spatially flat gauge we take $\psi=0$ 
and therefore $\gamma_{ij} = a^2 \delta_{ij}$. Furthermore, $\mathcal{N}$ is the lapse function while $\beta_{i}$ is the shift vector

In addition, there is the scalar perturbation $ \lambda=\lambda_{0}(t)+\delta \lambda (t,\vec{x})$ for the Lagrange multiplier. Substituting these perturbations back into the action (\ref{action1}) and solving for the constraint equations, we obtain the quadratic action for the field fluctuations $Q^{a}$.

\subsection{Background solutions}

In order to derive the background equations of motion, we expand the action (\ref{action1}) up to the linear order of scalar perturbations (\ref{pertubation}) and (\ref{eq:mapping2}) by performing the analysis in spatially flat gauge. 
Substituting these solutions back into the action (\ref{action1}), the first order action becomes
\begin{eqnarray}\label{actionv1}
\nonumber  S_{1}=\mathcal{V} \int {\rm d}t \ a^3 \Big[\Big(G_{ab}(&6& H \lambda_{0}+2 \dot{\lambda_{0}})\dot \phi_{0}^{a}-V_{b}+2 G_{ab}\lambda_{0} D_{t} \dot \phi_{0}^{a}\Big) Q^{b}+ (-V+3 M_P^2 H^2\\
&+&  \lambda_{0}(1+G_{bc}\dot{\phi}_{0}^{b}\dot{\phi}_{0}^{c} )\alpha+ \delta \lambda \Big(1- G_{bc}\dot{\phi}_{0}^{b}\dot{\phi}_{0}^{c})\Big) \Big] 
\end{eqnarray}
where $V^{b}\equiv G^{ab} \partial_{b} V$ and $\mathcal{V}=\int {\rm d}^{3}x$ is the spatial volume which will be assumed to be large enough but finite. 

Clearly, the variation with respect to Lagrangian multiplier perturbation mode $\delta \lambda$ gives us the mimetic constraint at the background level, i.e.,
\begin{equation}\label{mim1}
G_{ab} \dot \phi_{0}^{a}\dot \phi_{0}^{b}=1 \hspace{0.5cm} \text{or} \hspace{0.5cm} 
 G_{ab} D_{t} \dot\phi_{0}^{a}\, \dot\phi_{0}^{b}=0 \,,
\end{equation}
 in which the convenient derivative is given by 
\begin{align}
\label{D_t}
D_{t} X^{a}=\dot X^{a}+\Gamma_{bc}^{a} X^{b} \dot \phi_{0}^{c} \,.
\end{align} 
Here we have used $D_{t}G_{ab}=0$ which follows from the definition of the covariant differentiation. 

Now, we can derive additional equations of motion by varying with respect to the field fluctuation $Q^{a}$ and $\alpha$. Taking the variation of (\ref{actionv1}) with respect to $Q^{b}$, we arrive at
\begin{equation}\label{EOM3}
 \lambda_{0} D_{t} \dot \phi_{0}^{a}=\frac{V^{a}}{2}-(3 H \lambda_{0}+\dot{\lambda_{0}}) \dot \phi_{0}^{a} \, ,
\end{equation} 
in which $H=\dot{a}(t)/a(t)$ is the Hubble expansion rate.

\begin{figure}[t!]
	\begin{center}
		{\includegraphics[width=0.49\textwidth]{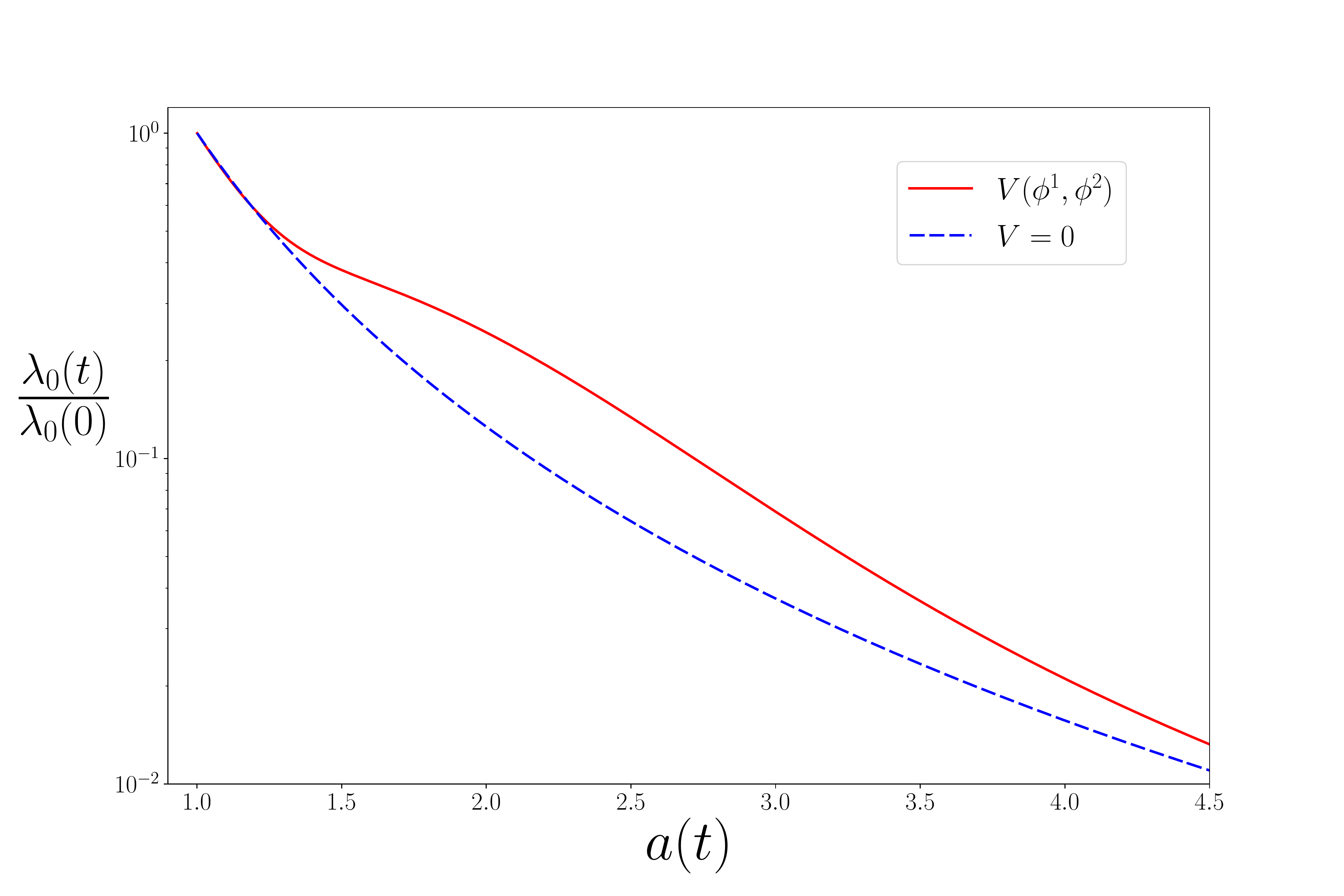}} 
		\includegraphics[width=0.49\textwidth]{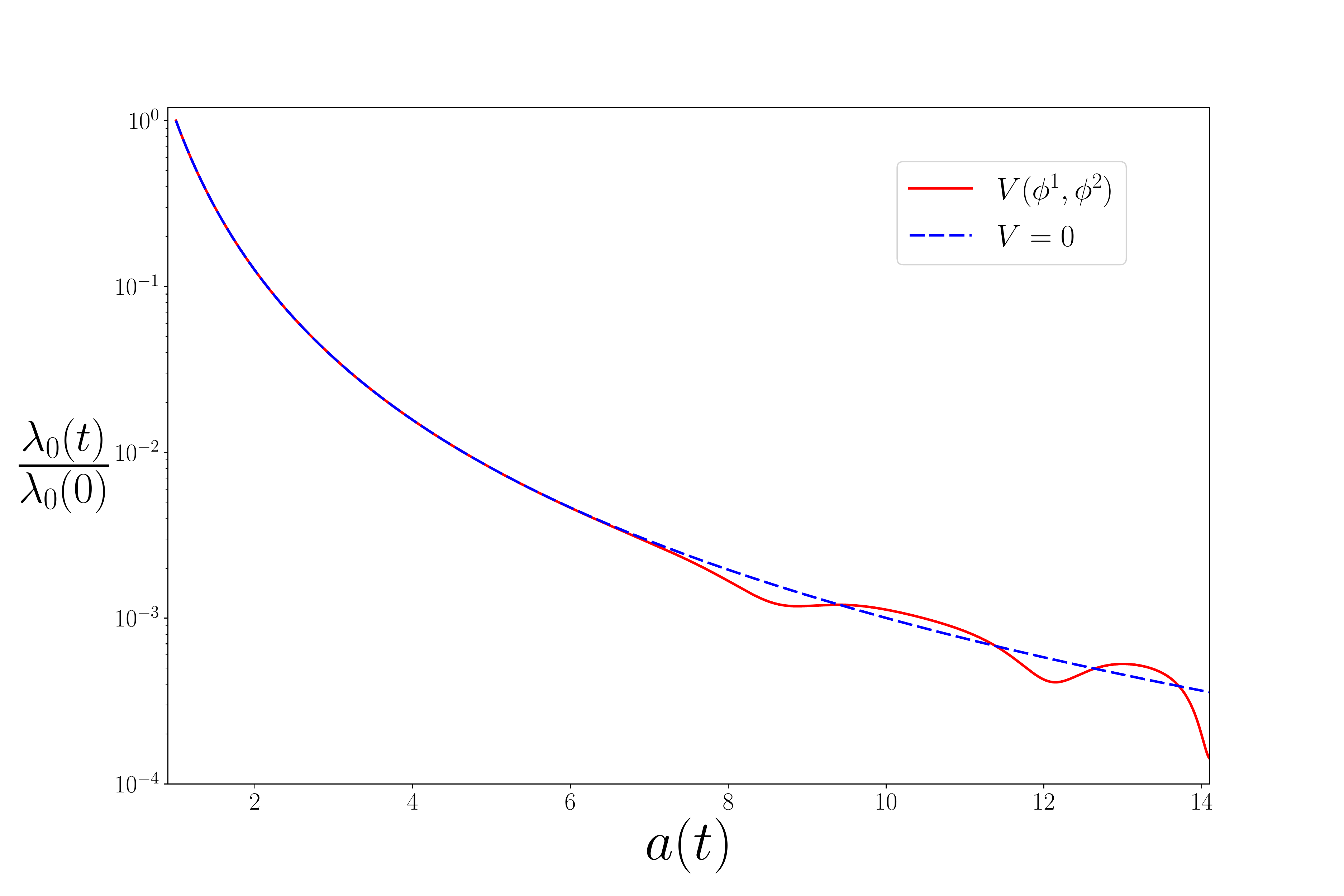}
		\caption{\footnotesize The evolution of the Lagrange multiplier for two different cases $V \neq 0$ (red solid curve) and $V=0$ (blue dashed curve). \textit{\textit{Left panel}}: The axion inflation example with the metric (\ref{G_Alen}) and potential (\ref{V_Alen}). \textit{Right panel}: The orbital inflation example with the metric (\ref{G-orbit}) and potential (\ref{p-orbit}). The $V=0$ case corresponds to a matter dominated Universe in which $\lambda(t) \propto a(t)^{-3}$ or equivalently $\rho \propto a^{-3}$ and $H=2/(3t)$. 
}
		\label{fig-Ht}
	\end{center}
\end{figure}

After contracting both sides of the above equation by $\dot{\phi}_{0}^{b}$ and using Eq. (\ref{mim1}), we obtain
\begin{equation}\label{EOM1}
\dot{\lambda_{0}}=\frac{1}{2} \Big(V_{a} \dot \phi_{0}^{a}-6 H  \lambda_{0}\Big) \,.
\end{equation}
The other equation comes from the variation of the action with respect to $\alpha$, yielding,
\begin{equation}\label{EOM2}
3 M_P^2 H^2+2  \lambda_{0}=V \,.
\end{equation} 
As we take time derivative from both sides of the above equation and compare with Eq. \eqref{EOM1}, we can immediately find
\begin{equation}\label{eqqq1}
\lambda_{0}=M_P^{2} \dot H \,.
\end{equation}  
Now combining  Eqs. (\ref{eqqq1}) and (\ref{EOM2}) one obtains the Friedmann equation,
\begin{equation}\label{eqqqQ1}
M_P^2(3 H^2+2 \dot H)=V \,,
\end{equation}
In the absence of the potential function, using Eqs. \eqref{eqqq1}, \eqref{eqqqQ1} and \eqref{EOM1}, one finds $\lambda_{0} \sim 1/a^3$ and $H(t)=2/(3t)$ (see Fig.~\ref{fig-Ht} for an example).  

On the other hand, by varying the action \eqref{action1} with respect to the metric $g_{\mu \nu}$, the effective energy momentum tensor is given by
\begin{equation}
T^{\mu}_{\nu}=-V \delta^{\mu}_{\nu}-2 \lambda G_{ab}\partial^{\mu} \phi^{a}\partial_{\nu} \phi^{b} \,.
\end{equation}   
Correspondingly,  one can read the energy density and the pressure as $\rho=T_{0}^{0}$ and $P=\frac{1}{3}T_{i}^{i}$.

For the case with no potential, $V=0$,  using the mimetic constraint \eqref{mim1} and $\lambda_{0} \sim 1/a^{3}$ we obtain $\rho=2 \lambda_{0} \propto a^{-3}$ and $P=0$. Therefore, similar to the standard single field mimetic theory \cite{Chamseddine:2013kea}, the multi-field generalization of the setup mimics  the role of dark matter.\footnote{We comment that by expanding the mimetic cosmology to models containing multiple gauge fields with a certain mimetic constraint imposed on the gauge field  strength tensor, the Maxwell term can play the role of the cosmological constant yielding to a  de Sitter-like spacetime  \cite{Gorji:2018okn,Gorji:2019ttx}.}

In order to understand the effect of the mimetic constraint \eqref{mim1} on the fields trajectory in the curved field space, let us consider, for example, an interesting two-field model proposed in  \cite{McDonough:2020gmn} for axion inflation. In this model, the axion field $\phi^{2}$ is characterized as the phase of a complex scalar field $\Phi$ ($\Phi=\phi^{1} e^{i \phi^{2}}$) in a canonical $U(1)$ symmetry-breaking model in which  the complex scalar field $\Phi$ has a non-minimal coupling to gravity in the Jordan frame, i.e. $f(\Phi)R$. Moreover, the radial
component $\phi^{1}$, as a second scalar field, plays the role of the order-parameter
of the symmetry breaking. The non-minimal coupling  depends only on the radial field, i.e. $f= f(\phi^1)$, and the kinetic terms of these  polar 
coordinates have the flat field space metric  in Jordan frame.

{On the other hand, going to the Einstein frame by performing a conformal transformation of the
spacetime metric such as $g_{\mu \nu}=2f(\phi^{1})/M_{p}^2 \tilde{g}_{\mu \nu}$, the field-space metric becomes \cite{McDonough:2020gmn},
\begin{align}
	\label{G_Alen}
	G_{11}=\dfrac{M_{P}^2}{2f} \left(
	1+\dfrac{3f_{,\phi^{1}}^2}{f}
	\right)
	\,,
	\hspace{0.5cm}
	G_{22}=\dfrac{M_{P}^2}{2f}(\phi^1)^2, \hspace{0.5cm} G_{12}=G_{21}=0 \,,
\end{align}
which describes a \textit{curved}  field space with the coupling $f$ given by $f(\phi^{1})=\frac{1}{2}\left(M^2+\xi (\phi^1)^2\right)$. Moreover, the potential function takes the following form in the Einstein frame \cite{McDonough:2020gmn}
\begin{align}
	\label{V_Alen}
	V(\phi^{1},\phi^2)=\dfrac{\zeta M_{P}^4}{16}\dfrac{((\phi^1)^2-v^2)^2}{f^2}+\dfrac{M_{P}^4\Lambda^4}{4f^2}(1-\cos(\phi^2)) \,.
\end{align}
Here $M,\, \xi,\, \epsilon_V,\, v \, {\rm and}\,\,\Lambda$ are parameters of the 
model. As seen, the potential function consists of two terms. The first term is related to the spontaneous symmetry breaking (Higgs) potential of the radial field and the second term comes from the nonperturbative effects which generate
a potential for the axion field.}

{As another example, we consider a two-field model with a polar parametrization, $\phi^{a}=(\phi^{1}, \phi^2)$, with a shift symmetry through the angular direction  $\phi^{2} \to \phi^{2}+c$~\cite{Achucarro:2019mea}. In this respect, the radial field space coordinate $\phi^1$ is orthogonal to the isometry while the angular coordinate $\phi^2$ is tangential to it. The simple form of a such field space in ``\textit{orbital inflation}" model \cite{Achucarro:2019mea} is defined as follows  
\begin{equation}\label{G-orbit}
G_{11}=1 \, , \hspace{0.5cm} G_{22}=f(\phi^{1})\,, \hspace{0.5cm} G_{12}=G_{21}=0\,,
\end{equation}  
in which one chooses $f(\phi^{1})=e^{2 \phi^{1}/R_{0}}$ associated with a hyperbolic field metric which has the Ricci curvature $\mathbb{R}=-2/R_{0}^2$. In addition, for a specific model of the orbital inflation, the potential function takes  the following form, 
\begin{eqnarray}\label{p-orbit}
V(\phi^{1},\phi^{2})&=&3 \mathcal{A} \Big(\phi^2-\frac{2}{3 f}\Big)\Big(1+\frac{\beta}{2} (\phi^1-\varphi)^2+\dfrac{\alpha}{6} (\phi^1-\varphi)^3\Big)^2\\
\nonumber &-& 2 \mathcal{A}^2 (\phi^2)^2 \Big(\beta (\phi^1-\varphi)+\frac{\alpha}{2}(\phi^1-\varphi)^2\Big)^2 \,.
\end{eqnarray} 
Notice that inflation happens at constant radius $\phi^{1}=\varphi$ in the orbital inflation setup, but here we only consider this model as a toy model with free parameters $\mathcal{A}$, $\alpha$, $\beta$, $R_{0}$, and $\varphi$.}

\begin{figure}
	\centering
	{\includegraphics[width=0.49\textwidth]{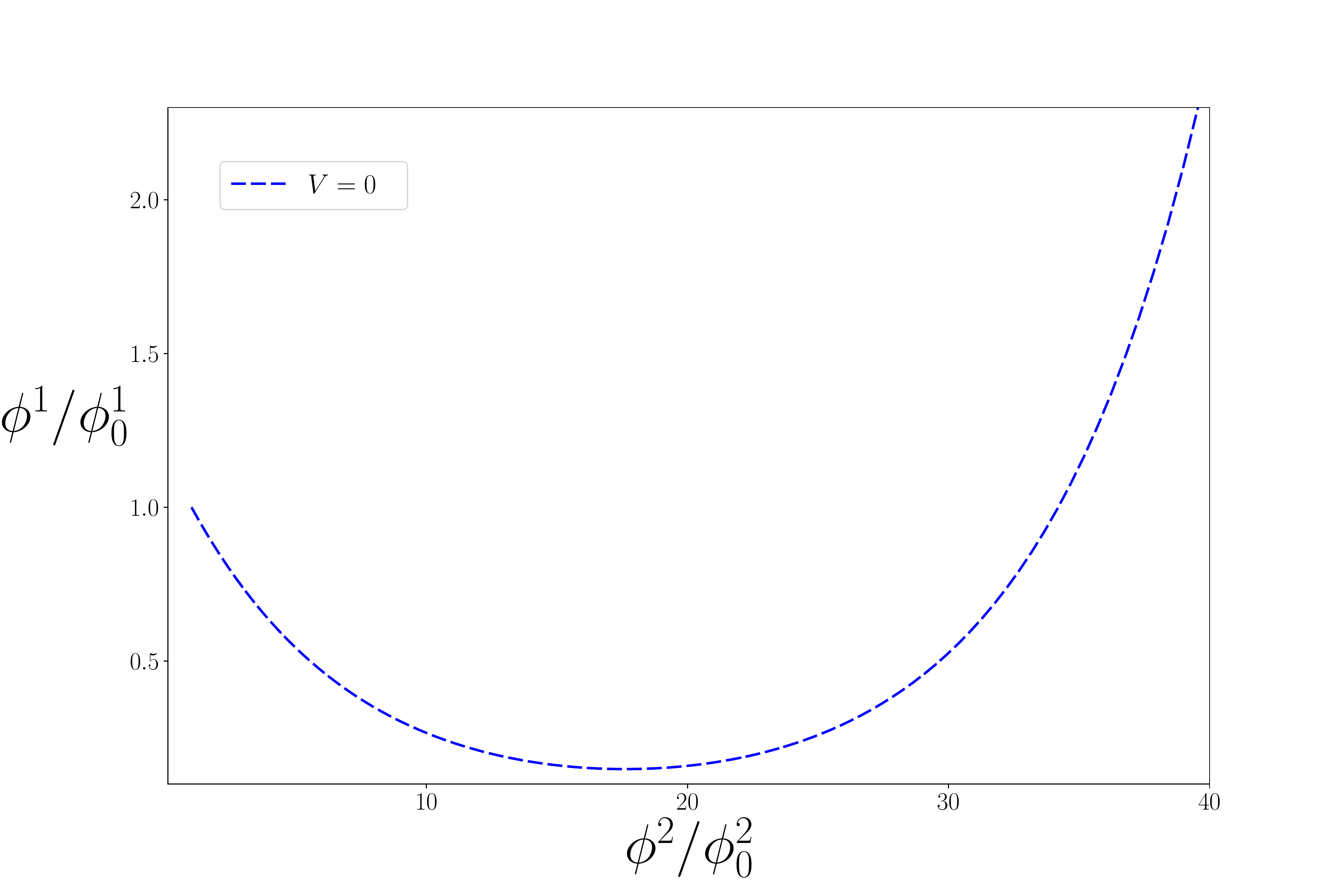}}
	{\includegraphics[width=0.49\textwidth]{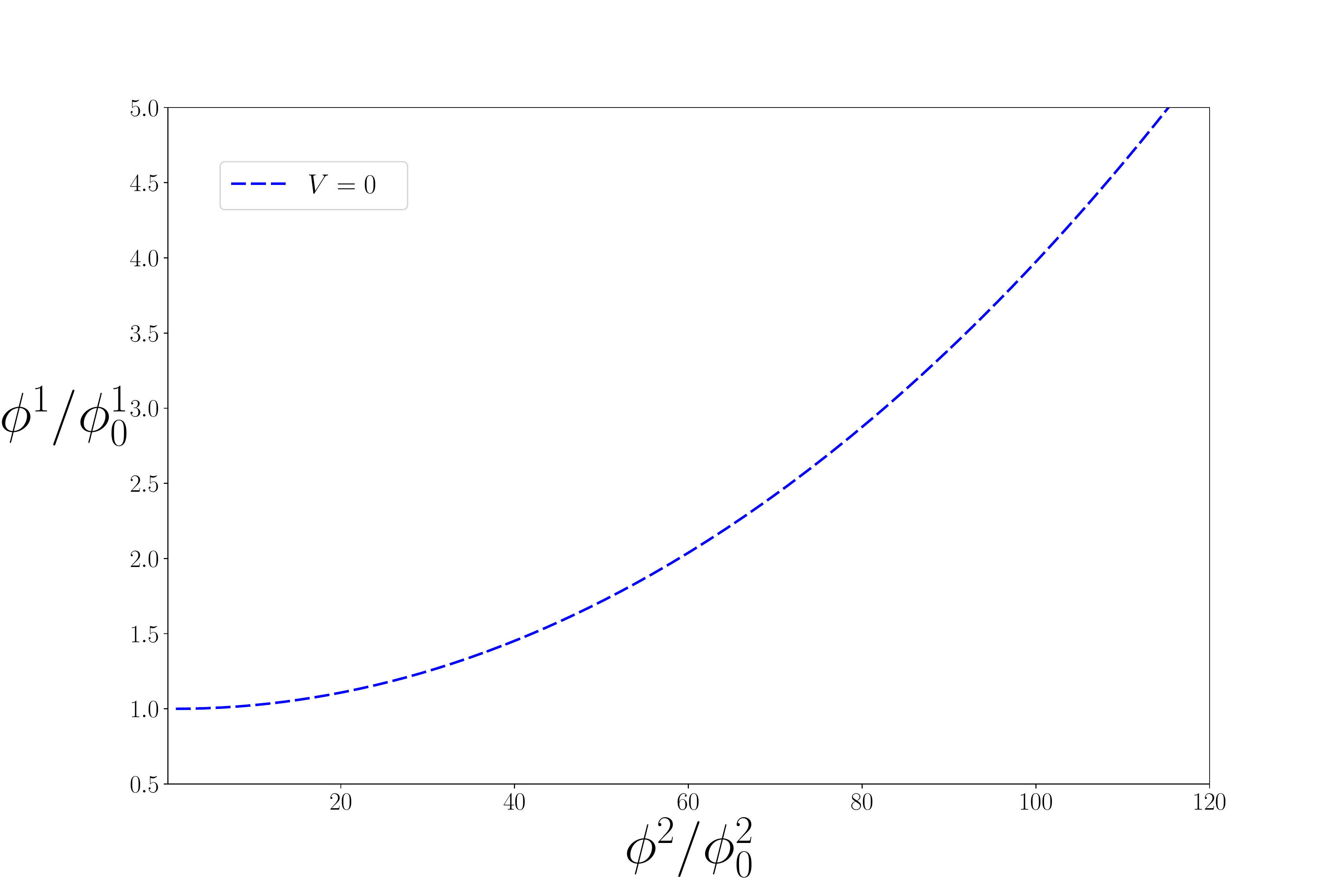}} 
	{\includegraphics[width=0.49\textwidth]{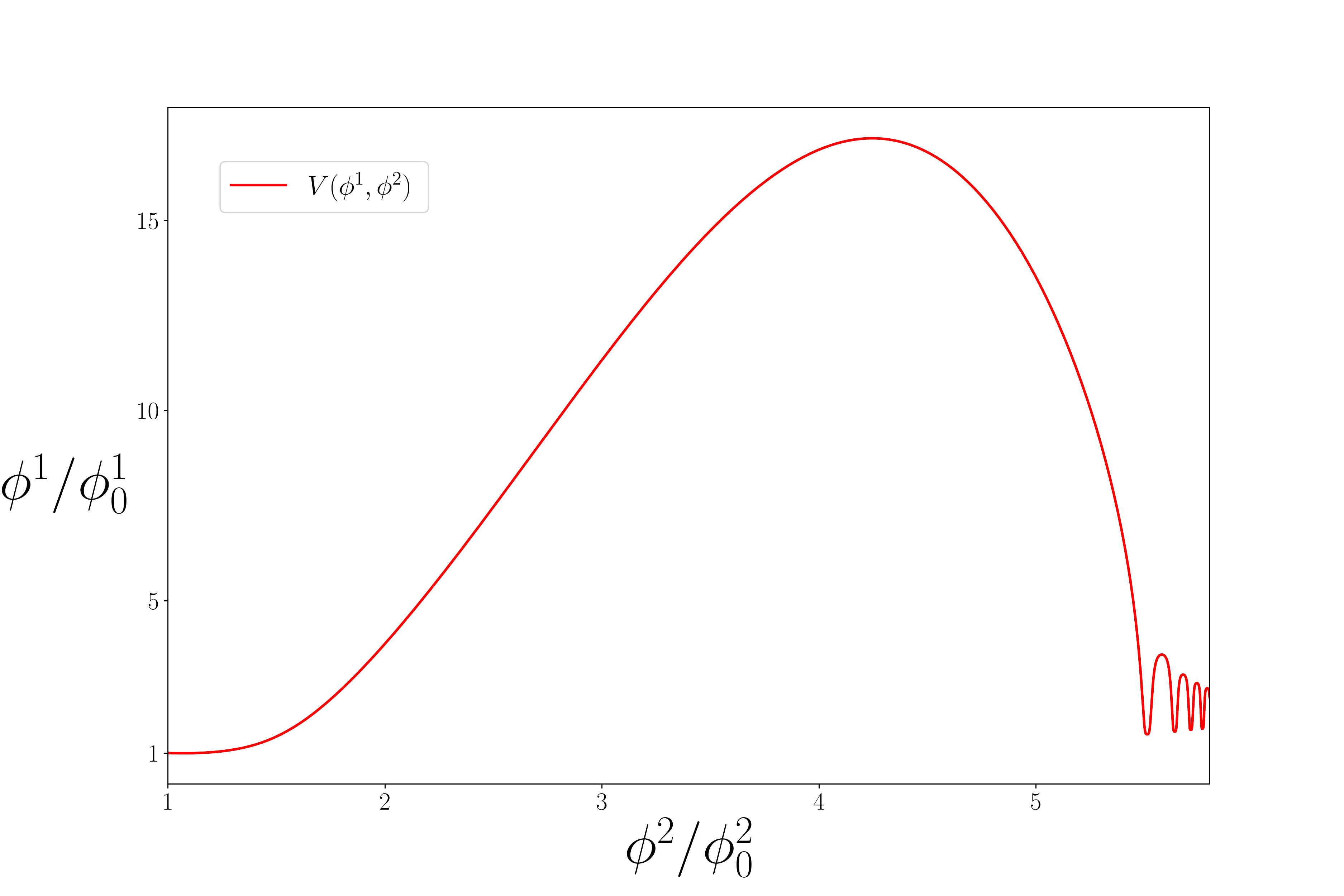}}
	{\includegraphics[width=0.49\textwidth]{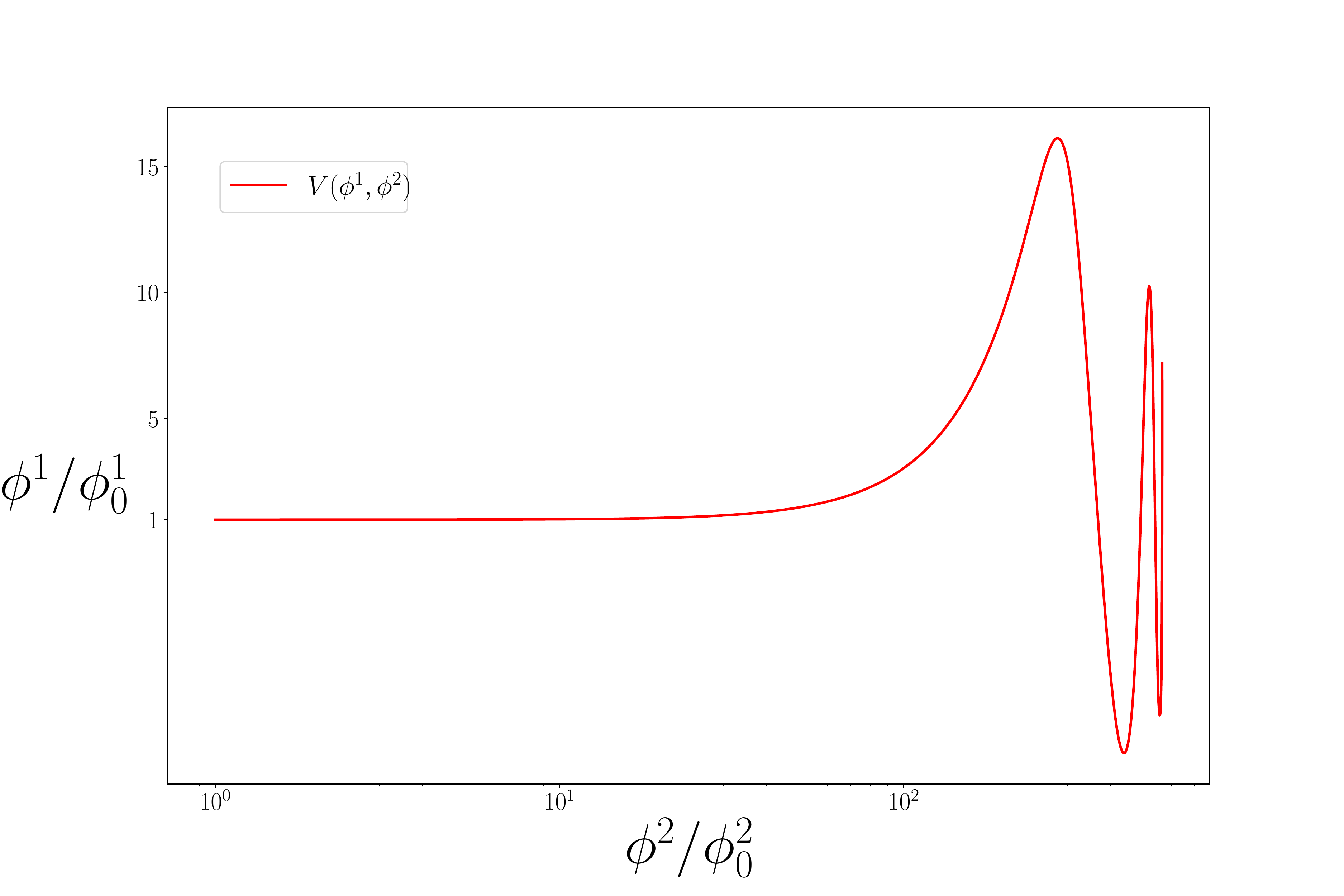}} 
	{\includegraphics[width=0.42\textwidth]{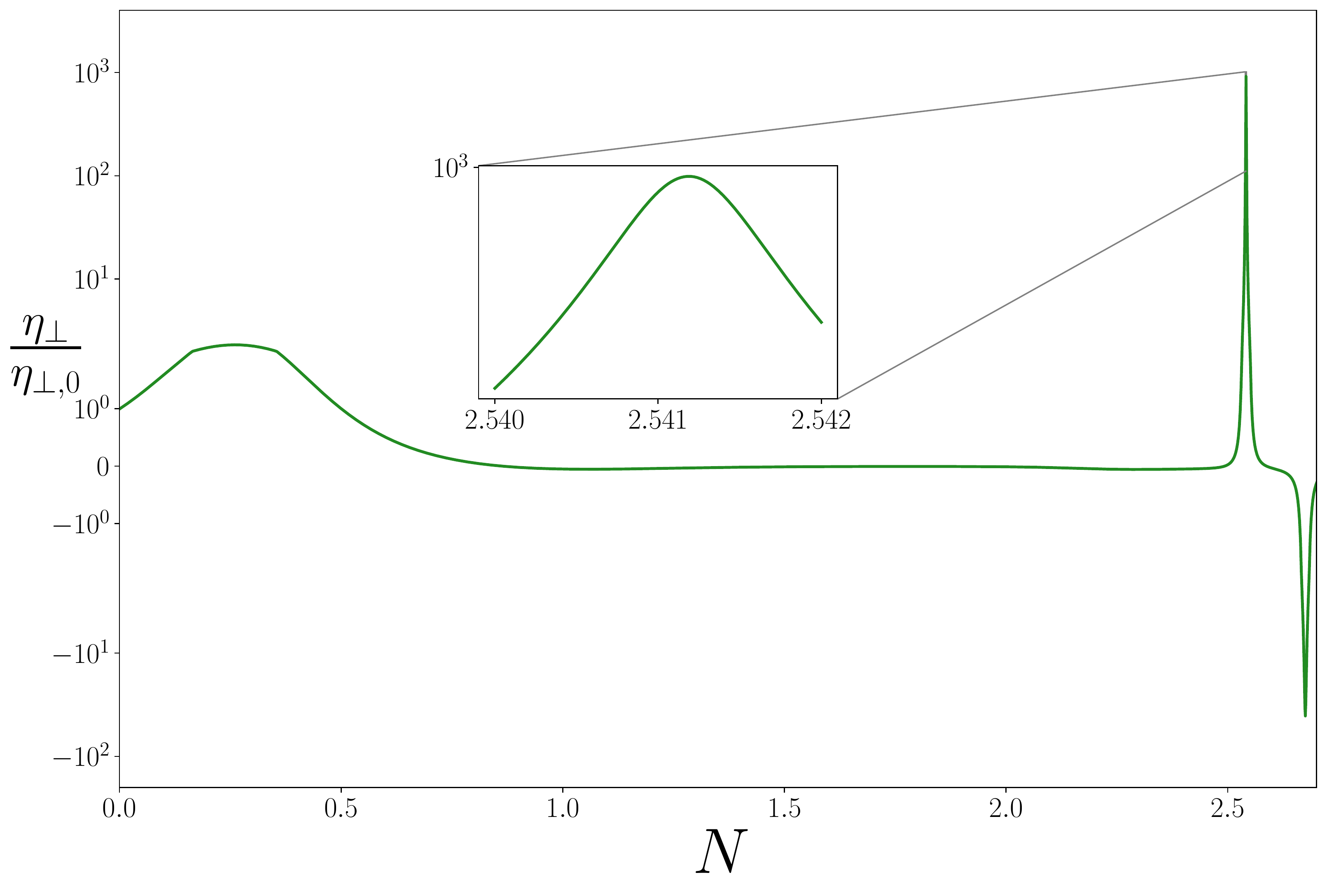}}
	\hspace{.8cm}
	{\includegraphics[width=0.42\textwidth]{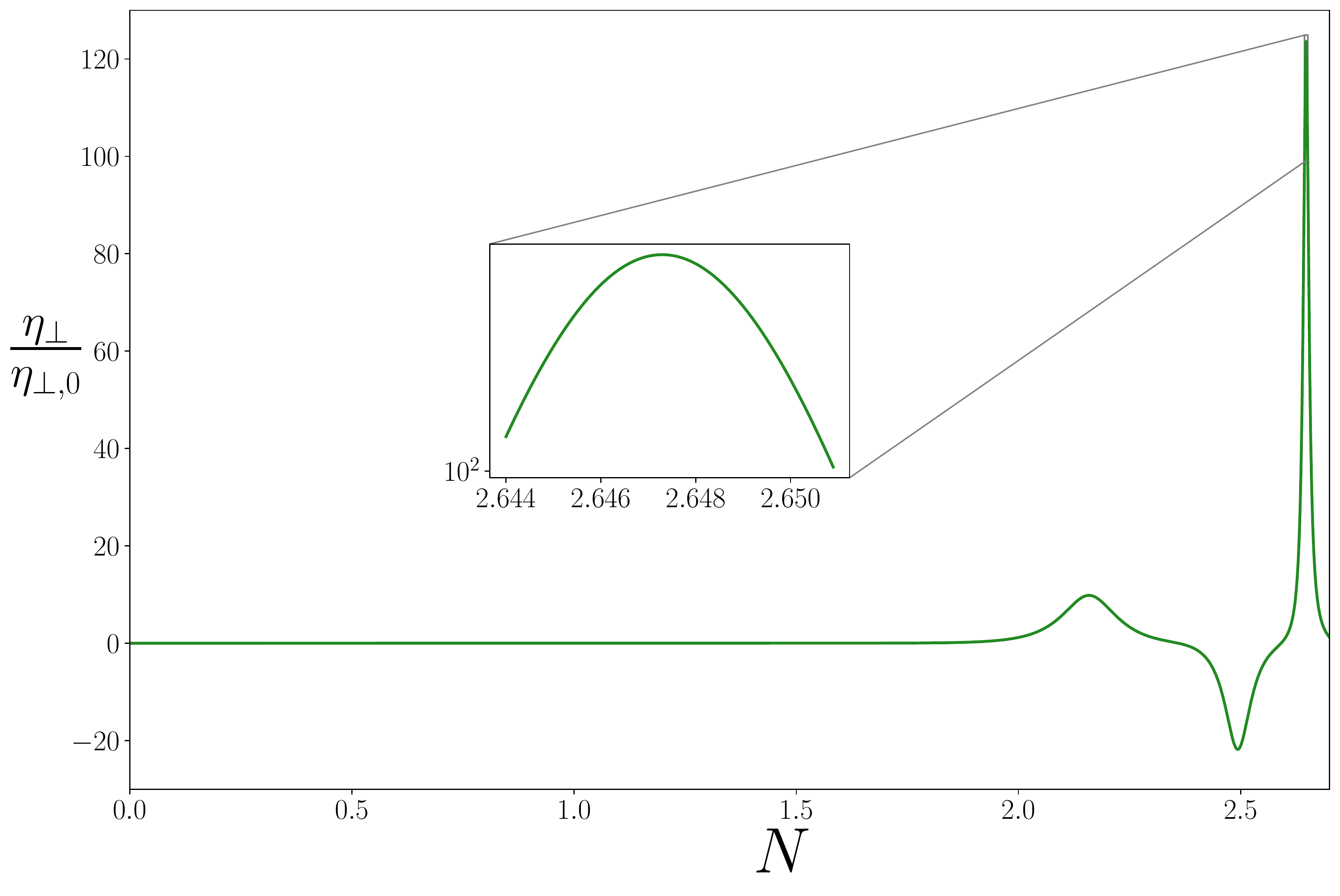}} 
	\caption{\footnotesize Illustration of the trajectory in field space and the evolution of the turning rate $\eta_{\perp}$. \textit{\textit{Left panels}}: The axion inflation example with the metric (\ref{G_Alen}) and potential (\ref{V_Alen}). \textit{Right panels}: The orbital inflation example with the metric (\ref{G-orbit}) and potential (\ref{p-orbit}). The parameters are  the same as those used in Fig.~\ref{fig-Ht}. The oscillatory behaviour of the trajectory (or changing sign in the tuning rate $\eta_{\perp}$)
		is due to the centrifugal force of the potential. This behaviour can also be seen from the turning rate parameter $\eta_{\perp}$ plotted in Fig.~\ref{fig-eta}. Although the trajectory for the case $V=0$ is curved, the rate of the turn is zero.}
	\label{fig-phi1phi1}
\end{figure}

In Fig.~\ref{fig-Ht}, we have presented the evolution of $\lambda_0(t)$ in the field space for the above two metrics for both cases $V=0$ and $V\neq0$. The \textit{left panel}  refers to the axion inflation example with the metric (\ref{G_Alen}), while the \textit{right panel} corresponds to the case of orbital inflation with the metric (\ref{G-orbit}) and potential (\ref{p-orbit}).   The red solid and blue dashed curves represent  
the cases $V=0$ and $V \neq 0$ respectively. As we see for the case $V=0$, the curve for $\lambda_0$ (or equivalently $\rho$) scales like $1/a^3$ which describes the energy density in matter-dominated era, whereas one observes a deviation form $1/a^3$ when  $V\neq 0$. For the case of axion inflation, the numerical parameters are  $\phi^1_0=0.5\,, \phi^2_0=0.97\pi,\dot{\phi}^1_0=-0.025, M=10^{-2} \,, \lambda_{0}(0)=-0.375\,, \xi=1\,, v=\sqrt{\frac{99}{100}}\,,\zeta=0.07\,,\Lambda=2.74 \times 10^{-3}$ in units of $M_P$. 
In the case of orbital inflation, we choose  $\phi^1_0=0.5\,, \phi^2_0=0.1,\dot{\phi}^1_0=0, R_0=50 \,, \beta=0.1\,, \alpha=0.01\,, \varphi=1\,,{\cal A}=10^{-4}$.

In addition, Fig.~\ref{fig-phi1phi1} shows the background trajectory in the field space for the above two examples. As illustrated in first two rows of Fig.~\ref{fig-phi1phi1}, one can observe sudden turns in the trajectory where the potential function $V$ is non-zero.

\subsection{Quadratic action}

In this part, we present the cosmological perturbations analysis in the spatially flat gauge. By expanding the action (\ref{action1}) up to the second order of scalar perturbations (\ref{pertubation}) and \eqref{eq:mapping2}, the quadratic action takes the following form,
\begin{eqnarray}\label{actionv2}
\nonumber 	&S^{2}&=\int {\rm d}t \  {\rm d}^3 x \  a^3 \Big[M_P^2 \epsilon H^2 G_{bc}D_{t}Q^{b}D_{t} Q^{c}-\Big(\frac{1}{2} V_{bc}+M_P^2 \epsilon H^2 \mathbb{R}_{bdcf} \dot \phi_{0}^{d} \dot \phi_{0}^{f}\Big)Q^{b}Q^{c}\\
&+&2 \delta \lambda \Big(\alpha-D_{t}Q_{b} \dot{\phi}_{0}^{b}\Big)- \alpha \Big(V_{b}Q^{b}+3 M_P^2 H^2 \alpha-M_P^2 \epsilon H^2 \alpha+2 \epsilon H^2 D_{t}Q_{b} \dot \phi_{0}^{b} \Big)\\
\nonumber &-&2 M_P^2H\Big(\alpha-\epsilon H Q_{b} \dot \phi_{0}^{b}\Big) \partial_{i} \partial^{i} B
-\frac{M_P^2}{a^2} \epsilon H^2 G_{ab} \partial_{i}Q^{b}\partial_{i}Q^{c}\Big] \,,
\end{eqnarray}
where $V_{ab}=V_{;ab}$ and 
\ba
\mathbb{R}^{a}_{bd,c}\equiv \Gamma^{a}_{bd,c}-\Gamma^{a}_{bc,d} +\Gamma^{a}_{ce}\Gamma^{c}_{bd}-\Gamma^{a}_{de}\Gamma^{e}_{bc} \, ,
\ea 
is the Riemann tensor associated with the curved  field space, and $\epsilon$ is the ``slow-roll'' parameter which is defined by $\epsilon  \equiv  -\dot{H}/H^2$.

From the above quadratic action, we see that the quadratic Lagrangian is linear in terms of the non-dynamical mode $B$, from which its equation of motion yields
\begin{equation}
\label{alpha}
\alpha=\epsilon H Q_{b} \dot \phi_{0}^{b} \,.
\end{equation}
It is worth mentioning that the linearity of the $B$ mode in the quadratic action causes the 
adiabatic perturbation to be non-dynamical  in the two-field mimetic example (see  below).

Plugging the relation (\ref{alpha}) in action (\ref{actionv2}), the reduced action takes the following form
\begin{eqnarray}\label{action22}
\nonumber S^{2} &=& \int d^3 x dt \, a^3 \Big[M_P^2 \epsilon H^2 G_{bc}D_{t}Q^{b} D_{t}Q^{c} -2 M_P^2  \epsilon^2 H^3 Q_{b} D_{t} Q_{c} \dot \phi_{0}^{b} \dot \phi_{0}^{c}-M^{2}_{bc} Q^{b} Q^{c}\\
& -& \frac{M_P^2}{a^2} \epsilon H^2 G_{bc} \partial_{i}  Q^{b}\partial^{i} Q^{c}+ 2 \delta \lambda \Big(\epsilon H Q_{b} \dot \phi_{0}^{b}-D_{t} Q_{b} \dot \phi_{0}^{b}\Big)\Big] \,,
\end{eqnarray}
where the effective mass matrix is defined via 
\begin{eqnarray}\label{massmatrix}
M_{ab}\equiv \frac{V_{ab}}{2}+ \frac{1}{2}\epsilon H \Big( V_{a} \dot{\phi}_{0b}+V_{b} \dot{\phi}_{0a}\Big)+ M_P^2 \epsilon^2 H^4(3-\epsilon) \dot{\phi}_{0a}\dot{\phi}_{0b}+M_P^2 \epsilon H^2 \mathbb{R}_{acbd} \dot{\phi}_{0}^{c}\dot{\phi}_{0}^{d} \,.
\end{eqnarray}

Finally, taking the variation of the above action with respect to $\delta \lambda$ yields the following relation
\begin{equation}\label{mimi2}
D_{t}Q_{b} \dot{\phi}_{0}^{b}=\epsilon H Q_{b} \dot{\phi}_{0}^{b} \,.
\end{equation}

Our analysis so far was general, valid for any number of of fields. In the rest of this section we restrict ourselves to the case of two-dimensional field space. This is because we will work in a new coordinate in field space where the perturbations are decomposed into the parallel to the background trajectory and perpendicular to it. While this decomposition is valid for field space of any dimension, but its geometric visualization is more simple in the case of 2D field manifold.

\begin{figure}[t]
	\begin{center}
		\includegraphics[scale=.6]{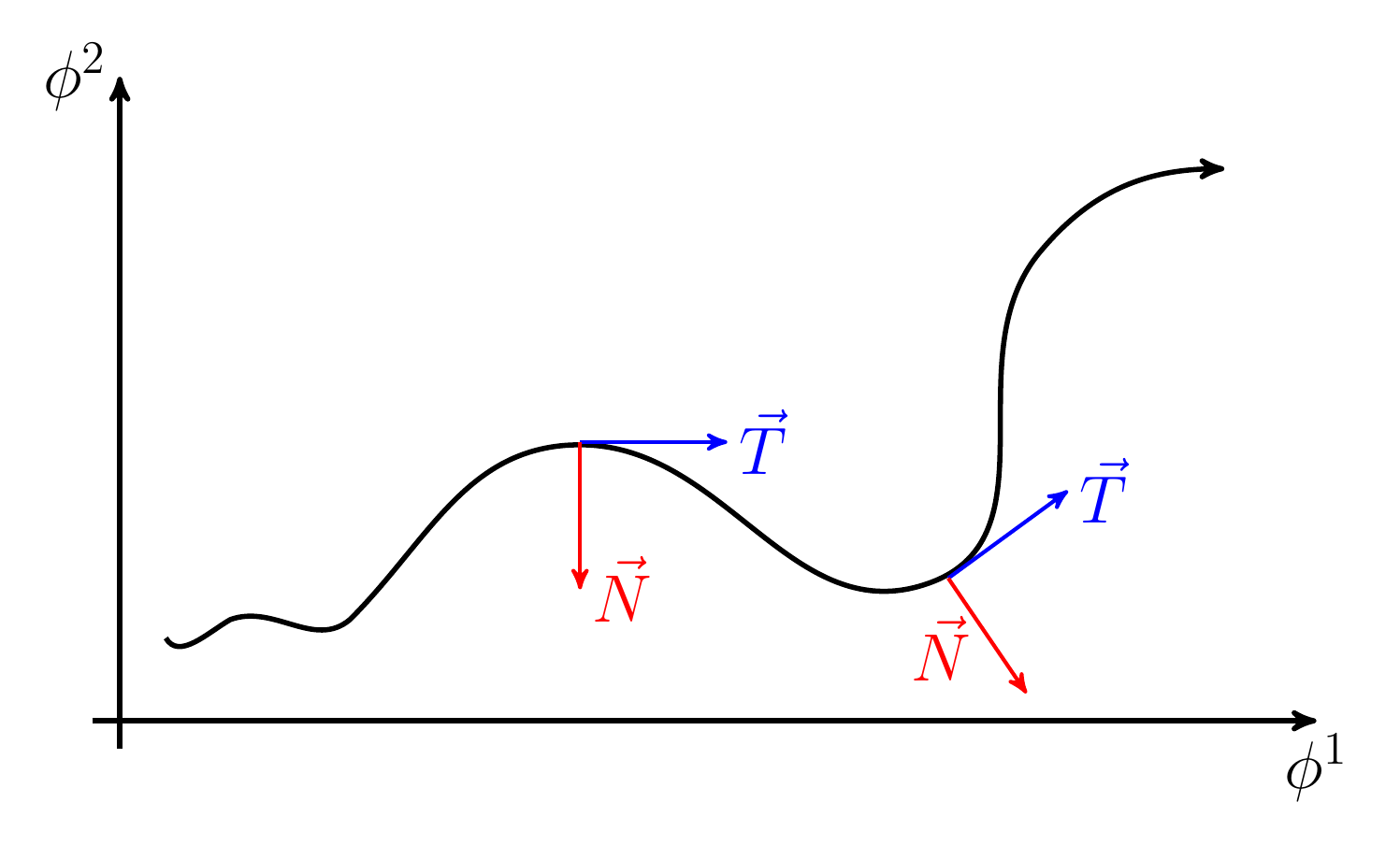}
		\caption{\footnotesize  A schematic view of the evolution of the tangent vector $T^a$ and normal vector $N^a$ along the field space trajectory. }
		\label{fig:trajectory}
	\end{center}
\end{figure}

In a 2D field space with the coordinate $\phi_0^a(t) = (\phi_0^1(t) , \phi_0^2(t))$, any trajectory defined in this space is parametrized by cosmic time $t$. To characterize this curve, it is useful to construct a set of orthogonal unit vectors $T^a$ and $N^a$ such that at a given time $t$, $T^a(t)$ is tangent to the path while $N^a(t)$ is normal to it \cite{Gong:2011uw,Elliston:2012ab}. In Fig.~\ref{fig:trajectory} we have illustrated a schematic plot of the evolution of the this set of orthonormal vectors along the background trajectory in the field space.  This set of vectors is defined as
\bea\label{eq1}
T^a &=& \dfrac{\dot \phi_0^a}{\dot{\phi_0}}  \\
N_a &=& \left( sgn(\pm 1) G \right)^{1/2} \epsilon_{a b} T^{b} ,
\eea
where $\dot{\phi_0^2}=G_{ab}\dot \phi_{0}^{a} \dot \phi_{0}^{b}$ which equals to one according to (\ref{mim1}). Moreover, $G$ is the determinant of the metric $G_{ab}$, the signum function $sgn(\pm 1)$ determines the signature of $G_{ab}$, for instance, $sgn(-1)$ is for Lorentzian signature, whereas $sgn(+1)$ is chosen for the Euclidean signature. In addition,  $\epsilon_{a b}$ is the two dimensional Levi-Civita symbol with $\epsilon_{11} = \epsilon_{22} = 0$ and $\epsilon_{12} = - \epsilon_{21} = 1$. 
These definitions satisfy the following conditions: $T_a T^a = 1$, $ N_a N^a=sgn(\pm 1)$ and $T^a N_a = 0$ \cite{Gong:2011uw,Elliston:2012ab}. Notice that the mimetic constraint (\ref{mim1}) in term of  this set of vectors can be written as $G_{ab}\dot \phi_{0}^{a} \dot \phi_{0}^{b}=T_{a}T^{a}=1$.

These two unit orthogonal vectors may be used to project the scalar field equation of motion  (\ref{EOM3}) along these directions. Projecting along $T^a$, we 
obtain Eq. (\ref{EOM2}) whereas projecting along $N^a$ one obtains the relation
\be
D_{t} \dot\phi_{0}^{a}=\frac{DT^a}{dt}=\frac{V_N}{2  \lambda_{0}}N^a, \label{time-deriv-T}
\ee
where $V_N \equiv N^a \partial_a V$. Clearly, this relation satisfies the mimetic constraint (\ref{mim1}). Now, let us define the second  ``slow-roll'' parameter $\eta^a$  as
\be
\eta^a  \equiv  -\frac{D\dot{\phi_0^a}}{dt}=-\eta_{\bot}N^a \, ,
\ee
with its normal component  given by $\eta_{\bot}  = V_N/2  \lambda_{0}$.
Combining this relation with (\ref{time-deriv-T}), we deduce the following relations,
\bea
\frac{DT^a}{dt} &=&   \eta_{\perp} N^a , \\
\frac{DN^a}{dt} &=& -\eta_{\perp} T^a .
\eea
\begin{figure}[t!]
	\begin{center}
		{\includegraphics[width=0.49\textwidth]{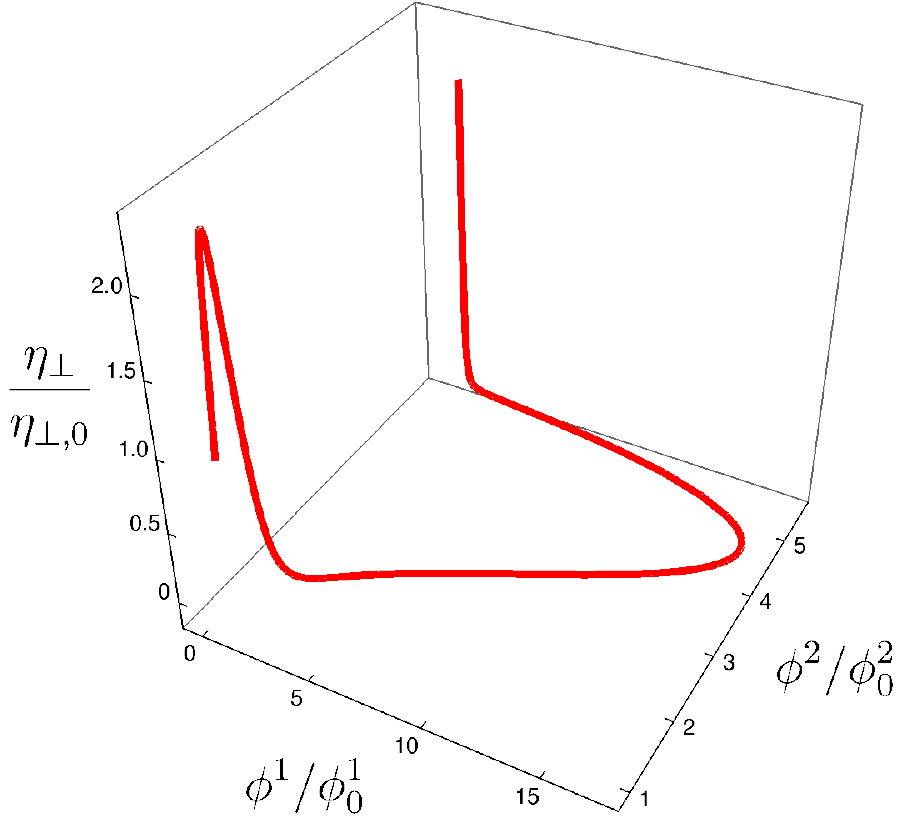}}
		{\includegraphics[width=0.49\textwidth]{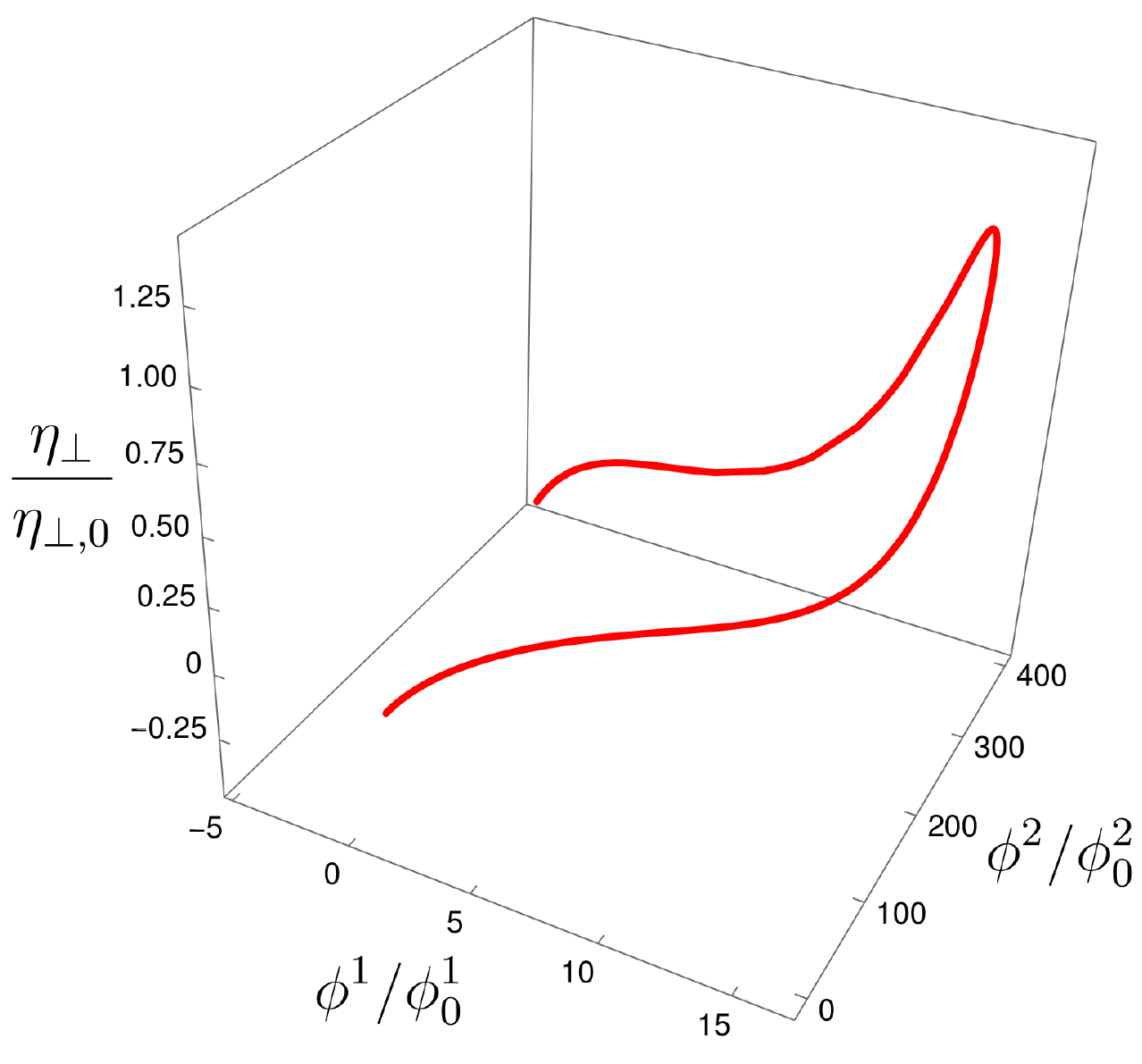}} 
		\caption{\footnotesize  The figure shows the evolution of turning rate $\eta_\perp$ in terms of the mimetic fields $\phi^1$ and $\phi^2$. \textit{\textit{Left panel}}: The axion inflation example with the metric (\ref{G_Alen}) and potential (\ref{V_Alen}). \textit{Right panel}: The orbital inflation example with the metric (\ref{G-orbit}) and potential (\ref{p-orbit}). The parameters are the same as those used in Fig.~\ref{fig-Ht}. This plot is consistent with Fig.~\ref{fig-phi1phi1} meaning that the turning toward left (right) in the field space indicates that $\eta_{\perp}$ increases (decreases). }
		\label{fig-eta}
	\end{center}
\end{figure}
Therefore, if $\eta_{\perp} = 0$ the vectors $T^a$ and $N^a$ remain covariantly constant with respect to $D_t$ (but not with respect to $\partial_t$, see the definition \eqref{D_t}) along the path. If  $\eta_{\perp} > 0$, the path turns to the left, whereas if $\eta_{\perp} < 0$ the turn is towards the right.
The value of $\eta_{\perp}$ therefore indicates how quickly the angle determining the orientation of $T^a$ varies in time.   Denoting this angle by $\theta$ we may therefore do the identification\footnote{Utilizing this definition,  we see that the ratio of curvature $\kappa_R$ characterizing the turning trajectory is defined as \cite{Gong:2011uw,Elliston:2012ab} $\kappa_R^{-1} \equiv  | \dot \theta |$.}
\be
\dot \theta \equiv  \eta_{\perp} . \label{angle-variation}
\ee
In the last row of Fig. \ref{fig-phi1phi1} and in  Fig.~\ref{fig-eta}, we have plotted the evolution of $\eta_{\perp}$ for the case $V \neq 0$ with respect to the number of efolds $N$ and fields $\phi^{1}$ and $\phi^2$, respectively.  As expected, the sudden turn in the field trajectory in Fig.~\ref{fig-phi1phi1} is translated into the change of the sign of $\eta_{\perp}$ in Fig.~\ref{fig-phi1phi1} and to moving to the left or right in Fig. \ref{fig-eta}.

For two-field case, the parallel and normal perturbations with respect to the background trajectory are given respectively by
\begin{align}
u^T & \equiv Q^T \equiv T_aQ^a \, ,
\\
u^N & \equiv Q^N \equiv N_aQ^a \, .
\end{align}
In this transformation $u^{T}$ corresponds to the perturbations parallel to the background trajectory while $u^{N}$ corresponds to perturbations normal to the trajectory which represents the entropy perturbations, i.e. the non-adiabatic mode \cite{Gordon:2000hv, Gong:2011uw,Elliston:2012ab}.  

By using the above representations and replacing $\dot \phi_{0}^{c}$ and $D_{t} \dot \phi_{0}^{c}$ by the tangent and normal vectors $T^{c}$ and $N^{c}$ according to Eqs. (\ref{eq1}) and (\ref{time-deriv-T}), the quadratic action (\ref{action22}), after straightforward
manipulations, reads
\begin{eqnarray}\label{qadraticaction1}
\nonumber S^{2}&=&\int d^3x dt a^3 M_P^2 \epsilon H^2\Big[ \mathcal{L}_{u_{N}}+\mathcal{L}_{u_{T}}+2sgn(\pm 1)\dot{\theta}u_{T}\dot{u}_{N}-2 \dot{\theta}u_{N}\dot{u}_{T}-2\Big( \frac{M_{NT}^2}{M_P^2 \epsilon H^2}\\
&-&\epsilon H \dot{\theta}\Big)u_{T} u_{N}+\frac{2}{M_P^2 H}\delta \lambda \Big( u_{T}-\frac{\dot{u}_{T}}{\epsilon H}+\frac{1}{ \epsilon H} \dot{\theta}u_{N}\Big)\Big] \,,
\end{eqnarray}
with
\begin{equation}
\mathcal{L}_{u_{N}}\equiv sgn(\pm 1) \Big(\dot{u}_{N}^2-\frac{1}{a^2} (\partial u_{N})^2\Big)+\Big(\dot{\theta}^2-\frac{M_{NN}^2}{M_P^2 \epsilon H^2}\Big) u_{N}^2 \,,
\end{equation}
and
\begin{equation}
\mathcal{L}_{u_{T}}\equiv\dot{u}_{T}^2-\frac{1}{a^2} (\partial u_{T})^2-2 \epsilon H u_{T} \dot{u}_{T}+\Big(sgn(\pm 1)\dot{\theta}^2-\frac{M_{TT}^2}{M_P^2 \epsilon H^2}\Big) u_{T}^2 \,,
\end{equation}
where $\eta_{H} \equiv \dot{\epsilon}/H \epsilon$ and the symmetric matrix $M_{IJ}$ elements are given by
\begin{eqnarray}
M_{NN}^2&\equiv& N^{a}N^{b}M_{ab}^2=\frac{1}{2} \Big(V_{NN}-sgn(\pm 1) \epsilon H^2 \mathbb{R}\Big) \,,
\\
M_{TT}^2&\equiv& T^{a}T^{b}M_{ab}^2=\frac{V_{TT}}{2}+\epsilon H \Big(V_{T}  +M_P^2 \epsilon H^3(3-\epsilon)\Big)
\,,
\\
M_{NT}^2&\equiv&M_{TN}^2= T^{a}N^{b}M_{ab}^2=\frac{1}{2} \Big(V_{NT}+ \epsilon H V_{N}\Big) \,,
\end{eqnarray} 
where $V_{NT}\equiv N^{a}T^{b}V_{;ab}$, $V_{NN}\equiv N^{a}N^{b}V_{;ab}$, and $V_{TT}\equiv T^{a}T^{b}V_{;ab}$. In particular, note the effect of the Ricci scalar $\mathbb{R}$ associated with the field space  manifold. Since we consider a 2D field space here, the Riemann tensor  can be expressed in the terms of the Ricci scalar $\mathbb{R}$ as
\begin{equation}
	\mathbb{R}_{abcd}=\frac{1}{2} \mathbb{R} \Big(G_{ac}G_{bd}-G_{ad}G_{cb}\Big).
\end{equation}

Finally, the equation of motion for $\delta \lambda$ then yields
\begin{equation}
	\dot{u}_{T}=\epsilon H u_{T}+\dot{\theta} u_{N} \, .
\end{equation}

Substituting the above result into Eq. \eqref{qadraticaction1},  we 
arrive at our final reduced quadratic action 
\begin{eqnarray}
\nonumber 	S^{2}&=&\int d^3 x dt a^3 M_P^2 \epsilon H^2 \Big\{sgn(\pm 1)\Big(\dot{u}_{N}^2-\frac{1}{a^2} (\partial u_{N})^2\Big)-\frac{M_{NN}^2}{M_P^2 \epsilon H^2} u_{N}^2
-\frac{1}{a^2} (\partial u_{T})^2\\
&+&\Big(sgn(\pm 1)\dot{\theta}^2-\epsilon^2 H^2-\frac{M_{TT}^2}{M_P^2 \epsilon H^2}\Big) u_{T}^2+2 sgn(\pm 1) \dot{\theta} u_{T} \dot{u}_{N}-2 \frac{M_{NT}^2}{\epsilon H^2}u_{N}u_{T}\Big\}
\end{eqnarray}

From the above action, it is clear that the perturbation mode perpendicular to background trajectory, $u_{N}$, is excited while the perturbation mode tangential to background trajectory, $u_{T}$, does not propagate. In other words, the entropy mode propagates with the speed of unity, whereas the sound speed for the adiabatic mode is zero. This is consistent with the fact that the mimetic background describes a fluid with no pressure. In addition, whether the entropy perturbation is free from the gradient as well as ghost instabilities directly depends on the signature of the metric, i.e. $sgn(\pm 1)$. In the case of the field space with an Euclidean signature ($sgn(+1)=1$), the entropy mode is healthy, whereas in the case of the Lorentzian manifold with $sgn(-1)=-1$, the entropy perturbation is pathological.   As one turns off the potential function $V$ in our setup and consider the flat metric, i.e, $G_{ab}=\delta_{ab}$ with an Euclidean signature, this action converts to Eq. (53) in \cite{Firouzjahi:2018xob} with the healthy entropy perturbation. However, when one considers $G_{ab}=\text{diag}(1,-1)$, the entropy mode always suffers from the ghost and gradient instabilities \cite{Shen:2019nyp}. In Append. \ref{AppB} 
we confirm these conclusions in the comoving gauge as well.  
 
Now let us check the number of degrees of freedom (DOFs) of the setup at the linear cosmological perturbation level by performing the Hamiltonian analysis of the quadratic action \eqref{qadraticaction1}. It is evident that the time derivative of $\delta \lambda$ does not appear in the action \eqref{qadraticaction1} so  one has the primary constraint $\Xi_{1}\equiv \Pi_{\delta \lambda}=0$ where $\Pi_{\delta \lambda}$ is the conjugate momentum associated with $\delta \lambda$.  After imposing this primary constraint, the total Hamiltonian is given by
\begin{equation}
H_{T}=\int d^3 x (\Pi_{u_{T}}\dot{u}_{T}+\Pi_{u_{N}}\dot{u}_{N}-\mathcal{L}+v \Xi_{1})
\end{equation}     
where $v$ is the Lagrangian multiplier. By constructing the conjugate momentum $\Pi_{u_{T}}=\partial \mathcal L/\partial  \dot{u}_{T}$ and $\Pi_{u_{N}}=\partial \mathcal L/\partial  \dot{u}_{N}$ from the quadratic action \eqref{qadraticaction1}, the total Hamiltonian in the Fourier space is written as
\begin{eqnarray}
\nonumber H_{T}&=&\int dx^3 \Big[\frac{\Pi_{u_{N}}^2}{4 sgn(\pm 1) a^3 M_P^2 \epsilon H^2}+\Big(sgn(\pm 1) a \epsilon H^2 k^2+a^3 M_{NN}^2\Big)u_{N}^2+ \frac{\Pi_{u_{T}}^2}{4  a^3 M_P^2 \epsilon H^2} \\
\nonumber &+&M_P^2 \epsilon H^2\Big(a k^2+a^3 \epsilon^2 H^2+a^3 \frac{M_{TT}^2}{\epsilon H^2}\Big)u_{T}^2+\frac{\delta \lambda}{M_P^2 \epsilon H^2}\Big(\Pi_{u_{T}}+a^3 \delta \lambda\Big) +2 a^3 M_{NT}^2 u_{N} u_{T}\\
&+& \epsilon H \Pi_{u_{T}} u_{T}+\dot{\theta} \Big(\Pi_{u_{T}} u_{N}-\Pi_{u_{N}} u_{T}\Big)+v \Xi_{1}\Big] \, .
\end{eqnarray}
Note that because we work in the Fourier space, all coordinate and momentum variables are only functions of time and the wave number $k$ in which, for simplicity, their dependence on $k$ are dropped.

In order to derive the secondary constraint, we should check the time evolution of the primary constraint $\Xi_{1}$ which amounts to examine the consistency relation. Beginning with $\Xi_{1}$, one obtains the following constraint,
\begin{equation}
\Xi_{2}\equiv \{\Xi_{1},H_{T}\}=-\frac{1}{M_P^2 \epsilon H^2}\Big(\Pi_{u_{T}}+2 a^3 \delta \lambda\Big) \approx 0 \, .
\end{equation}    
The above constraint can be solved for $\delta \lambda$, yielding
\begin{equation}
\delta \lambda= -\frac{\Pi_{u_{T}}}{2 a^3} \, .
\end{equation}

The next consistency relation gives
\begin{eqnarray}
\nonumber 	\Xi_{3}\equiv \{\Xi_{2},H_{T}\}&=&\frac{1 }{M_P^2 \epsilon H^2 }\Big(-2 a^3 v-\dot \theta \Pi_{u_{N}}+\epsilon \Pi_{u_{T}}+2 a^3 M_{NT}^2+2 a^3 M_{TT}^2\Big)\\
	&+& 2\Big( a k^2+a^3 \epsilon^2 H^2\Big) u_{T} \approx 0 \, ,
\end{eqnarray}
which determines the Lagrange multiplier $v$ as a function of the phase space variables and thus the chain of constraints for primary constraint $\Xi$ ends here. More precisely, both constraints $\Xi_{1}$ and $\Xi_{2}$ are second class constraints and therefore the physical number of DOFs is $(6-2)/2=2$. Together with two tensor modes, the total DOFs of our two-field setup is four. This is confirmed in next Section where we present the full 
non-linear Hamiltonian analysis. In spite of the fact that the adiabatic mode does not propagate at the linear perturbation level, it contributes to the physical DOFs of the model. The same result has been reported in \cite{Shen:2019nyp} for the flat two-field setup.

As seen, the adiabatic mode $u_{T}$ does not propagate in 2D field space  at linear order in perturbation.  This is similar to the simple single field mimetic model where the curvature perturbation was non-propagating at the linear order. To remedy this issue, in the case of single field mimetic setup,  higher derivative terms such as $(\Box \phi)^2$ were added to the theory yielding a non-zero sound speed for the curvature perturbations. Motivated by this fact, we may also consider adding higher derivative terms to our action. First consider the term 
\begin{equation}
Y=X_{ab}X^{ab}=G_{ad}G_{bc}X^{dc}X^{ab} \, ,
\end{equation} 
where $X^{ab}=\partial^{\mu}\phi^{a}\partial_{\mu}\phi^{b}$ \footnote{Note that the term $G_{ab}\Box \phi^{a} \Box \phi^{b}$ is not covariant in field space nor in spacetime.}.This term is linear with respect to the non-dynamical mode $ B$. In order to prove it let us write down $X^{ab}$ by using ADM variables as
\begin{align}
X^{ab} &= -\dfrac{1}{\mathcal N^2}
\left(
\dot{\phi}^a - \beta^{k}\partial_k \phi^a
\right)
\left(
\dot{\phi}^b - \beta^{k}\partial_k \phi^b
\right)
+ \gamma^{kl}\partial_k \phi^a \partial_l \phi^b \, .
\end{align}
By expanding the above relation up to the second order of scalar perturbations, we arrive at
\begin{align}
X^{ab} &= -\dot{\phi}^a_0 \dot{\phi}^b_0+2\alpha \dot{\phi}^a_0 \dot{\phi}^b_0-\left(\dot{\phi}^a_0 \dot{Q}^b+\dot{\phi}^b_0 \dot{Q}^a\right)-3\alpha^2 \dot{\phi}^a_0 \dot{\phi}^b_0+2\alpha\left(\dot{\phi}^a_0 \dot{Q}^b+\dot{\phi}^b_0 \dot{Q}^a\right)
\nonumber\\
&
-\dot{Q}^a\dot{Q}^b
+\partial^k B\left(
\partial_kQ^a \dot{\phi}^b_0+\partial_kQ^b \dot{\phi}^a_0
\right)+a^2\delta^{kl}\partial_k Q^a \partial_l Q^b  \, .
\end{align}
Clearly, the scalar $Y$ is linear with respect to the non-dynamical mode $ B$ up to the second order. This confirms that the adiabatic mode does not propagate in our model 
just by adding the $Y$ term to the action. Therefore, one may consider to couple non-minimally a function of  $Y$ term to the spacetime Ricci scalar $R$. Since the Ricci scalar is  linear with respect to $ B$, adding $f(Y)R$ to the action can not generate any non-linear contributions of $B$ as well. After eliminating non-dynamical modes in the corresponding quadratic action, Eq. (\ref{mimi2}) is still satisfied and it prevents  the adiabatic mode 
$u_{T}$ to propagate. 

Similar to the single field mimetic scenario, let us now consider box terms such as 
\begin{equation}
(\Box L )^2 
\end{equation}
where $L^2\equiv G_{ab} \phi^a \phi^b$ describing the length in the field space. This box term is defined as
\begin{align}
\Box L = \dfrac{1}{\sqrt{-g}}\partial_\mu \left(
\sqrt{-g} g^{\mu \nu} \partial_\nu L
\right)
\supseteq
\partial_k g^{k0}\dot{L} + \ddot{L} \, .
\end{align}
Correspondingly, its contribution to the quadratic action yields 
\begin{equation}
S^{2}	\supseteq \int dx^3 dt \left( \Box L\right)^2 \supseteq  \int d^3x dt \Big((\partial_k \partial^k B)^2 \dot{\bar{L}}^2 + \delta\ddot{L}^2\Big) \, ,
\end{equation}
where $\bar{L}=G_{ab} \phi^a_0 \phi^b_0$ and $\delta L=L-\bar{L}$.
It is obvious that the quadratic action contains not only the gradient term $(\partial^2B)^2$, but also the higher derivative term $\delta \ddot{L}$ which is the source of the 
Ostrogradsky ghost. As mentioned earlier, incorporating a non-linear term for the $B$ mode in the quadratic action modifies the relation \eqref{mimi2}, causing the adiabatic modes $u_{T}$ to propagate. However, this comes with the price  that both entropy and adiabatic modes develop Ostrogradsky-type ghost.  This kind of ghost can be removed by applying explicit combinations of higher derivative terms used in Horndeski theories \cite{horndeski1974second} and the higher
derivative interactions coupled to gravity \cite{BenAchour:2016fzp} as for example employed 
in single field mimetic setup \cite{Zheng:2017qfs,Hirano:2017zox,Gorji:2017cai}.

\section{Non-linear Hamiltonian analysis}
\label{AH}

In this section we present the full non-linear Hamiltonian analysis of the system to count 
the correct number of degrees of freedom (DOFs).  This is motivated in part from the fact that the adiabatic mode is not propagating at the linear order of equation of motion so one may wrongly conclude that there is only one scalar degree of freedom. Although this issue is addressed in previous section using the quadratic Hamiltonian analysis, but here we confirm it employing full non-linear perturbation analysis. 

We perform the non-linear Hamiltonian analysis using the Arnowitt-Deser-Misner (ADM) decomposition \cite{arnowitt2008republication}. In fact, the ADM decomposition is used to characterize the nature of gravity as a constrained system.  
In accord with the ADM decomposition, the metric components of spacetime take the following form.
\begin{eqnarray} && \label{metricinv}
g_{00}=-\mathcal{N}^{2}+\beta_{i}\beta^{i},\hspace{7mm} g_{0i}=\beta_{i},\hspace{7mm} g_{ij}=\gamma_{ij},\label{a8} \nonumber\\ &&
g^{00}=-\frac{1}{\mathcal{N}^{2}},\hspace{7mm} g^{0i}=\frac{\beta^{i}}{\mathcal{N}^{2}},\hspace{7mm}g^{ij}=\gamma^{ij}-\frac{\beta^{i}\beta^{j}}{\mathcal{N}^{2}},\label{a9}
\end{eqnarray}
where as before $\mathcal{N}$ is the lapse function and $\beta^{i}$ is the shift vector. The spatial component $\gamma_{ij}$ is defined as a metric on the three-dimensional spatial hypersurface  embedded in the full spacetime. 

In the ADM framework, the action (\ref{action1}) can be written as
\begin{equation}\label{eq:ADMaction}
S=S_{G}+S_{M}
\end{equation}
where $S_{G}$ is associated to the pure gravity part, i.e.,
\begin{equation}
\label{eq:actionGR}
S_{G} = \int d^4 x \sqrt{\gamma} \mathcal N   \frac{M_P^2}{2}  \left( R^{(3)}+K_{ij}K^{ij}-K^2 \right)  \, ,
\end{equation}
where $R^{(3)}$ is the curvature of three-dimensional spatial hypersurface, constructed from $\gamma_{ij}$.  Moreover, $K_{ij}$ is the extrinsic curvature defined as
\begin{equation}
\label{eq:extrinsiccurvature1}
K_{ij} = \frac{1}{2\mathcal N} \left( \partial_t\gamma_{ij} - \beta_{i;j} - \beta_{j;i} \right) \, , \hspace{0.5cm} K \equiv K^i{}_i \, ,
\end{equation} 
with the covariant derivative being calculated with respect to $\gamma_{ij}$. 

On the other hand, one can write the matter part of the action as  
\begin{eqnarray}
\label{eq:ADMactionmatter}
\nonumber S_{M} = \int d^4 x \sqrt{\gamma} \mathcal N  \Big\{ \lambda \Big[-\dfrac{G_{ab}}{\mathcal N ^2}
\big(
\partial_{t}\phi^a - \beta^{k}\partial_k \phi^a
\big)
\big(
\partial_{t}\phi^b - \beta^{l}\partial_l \phi^b
\big)
+G_{ab} \gamma^{kl}\partial_k \phi^a \partial_l \phi^b+1\Big]
-V  \Big\} \, .
\end{eqnarray}
Note that these actions do not depend upon the time derivative of $\mathcal N$, $\beta^{i}$, and $\lambda$. It means that these quantities are not  dynamical variables. Consequently the dynamical variables are $\gamma_{ij}$ and $\phi^{a}$. The momentum canonically conjugate to $\gamma_{ij}$ and $\phi^{a}$, respectively, are
\begin{eqnarray}
\Pi^{ij}&=&\frac{\delta S_{G}}{\delta \partial_{t} \gamma_{ij}}=\frac{M_P^2}{2}\sqrt{\gamma}(K^{ij}-\gamma^{ij}K)\label{c1}\\
\Pi_{a}&=&\frac{\delta S_{M}}{\delta \partial_{t} \phi^{a}}=-\frac{2\sqrt{\gamma}}{\mathcal{N}} \lambda G_{ab}(\partial_{t}\phi^{b}-\beta^{i}\partial_{i}\phi^{b}) \, .
\end{eqnarray}
Plugging the above expression of the conjugate momenta in  the 
action \eqref{eq:actionGR}, the total mimetic action can be expressed as
\begin{eqnarray} \label{actionHam}&&
S=\int d^{4}x \Big[ \Pi^{ij}\partial_{t}\gamma_{ij}+\Pi_{a}\partial_{t}\phi^{a}-\mathcal{N}(\mathcal{H}_{G}+\mathcal{H}_{M})-\beta^{i}({\mathcal{H}_{G}}_{i}+{\mathcal{H}_{M}}_{i})\Big],
\end{eqnarray} 
where
\begin{equation}
\mathcal{H}_{G}\equiv - \frac{M_P^2}{2}\sqrt{\gamma} \ R^{(3)}-\dfrac{2}{M_P^2 \sqrt{\gamma}}\left(\frac{1}{2}\Pi^{2}-\Pi^{ij}\Pi_{ij}\right),\hspace{0.5cm} \mathcal{H}^{i}_{G}\equiv-2\Big(\partial_{j} \Pi^{ij}+\Gamma_{jk}^{i} \Pi^{jk}\Big)
\end{equation}
and
\begin{eqnarray} 
\mathcal{H}_{M}\equiv-\frac{G^{ab}}{4\lambda\sqrt{\gamma}}\Pi_{a}\Pi_{b}-\lambda\sqrt{\gamma}(\gamma^{ij}G_{ab}\partial_{i}\phi^{a}\partial_{j}\phi^{b}+1)+\sqrt{\gamma} V(\phi^{a}), \hspace{0.5cm}
\mathcal{H}_{M}^{i}\equiv\Pi_{a}\partial^{i}\phi^{a} \, ,
\end{eqnarray}
with $\Pi\equiv \Pi^{i}_{i}$. 

Now, we have five primary constraints $(\Pi_{\mathcal{N}},\Pi_{i},\Pi_{\lambda} )\approx 0$\footnote{Following \cite{bojowald2010canonical}, the notation $\approx 0$ stands for the constraint equations. } which are associated with non-dynamical variables $(\mathcal{N},\beta^{i},\lambda)$, respectively.
By taking into account these primary constraints and the action \eqref{actionHam}, we can construct the total Hamiltonian function from the standard definition in \cite{bojowald2010canonical} as follows,
\begin{eqnarray} &&
H_{T}=\int d^{3}x  [\mathcal{N}(\mathcal{H}_{G}+\mathcal{H}_{M})+\beta^{i}({\mathcal{H}_{G}}_{i}+{\mathcal{H}_{M}}_{i})+v_{\mathcal{N}}\Pi_{\mathcal{N}}+v^{i}\Pi_{i}+v_{\lambda}\Pi_{\lambda}], 
\end{eqnarray}
where $ v_{\mathcal{N}} $,$ v^{i} $  and $ v_{\lambda} $ are Lagrange multipliers which enforce the primary constraints. To identify the secondary constraints, we should check the time evolution of the primary constraints using the Poisson brackets \cite{dirac1964lectures}. 
For gravity and matter parts, the Poisson bracket is defined as
\begin{align}\label{poissonb}
\nonumber \{\mathcal{X}, \mathcal{Y}\} &\equiv 
\int d^{3}x
\Bigg[\frac{\delta\mathcal{X} }{\delta \mathcal{N}(x)}\frac{\delta\mathcal{Y} }{\delta \Pi_{\mathcal{N}}(x)}-\frac{\delta\mathcal{Y} }{\delta \mathcal{N}(x)}\frac{\delta\mathcal{X} }{\delta \Pi_{\mathcal{N}}(x)} +\frac{\delta\mathcal{X} }{\delta \beta^{i}(x)}\frac{\delta\mathcal{Y} }{\delta \Pi_{i}(x)}-\frac{\delta\mathcal{Y} }{\delta  \beta^{i}(x)}\frac{\delta\mathcal{X} }{\delta \Pi_{i}(x)} 
\nonumber\\& \hspace{1.5cm}
+\frac{\delta\mathcal{X} }{\delta \lambda(x)}\frac{\delta\mathcal{Y} }{\delta \Pi_{\lambda}(x)}-\frac{\delta\mathcal{Y}}{\delta \lambda(x)}\frac{\delta\mathcal{X} }{\delta \Pi_{\lambda}(x)} +\frac{\delta\mathcal{X} }{\delta \gamma_{ij}(x)}\frac{\delta\mathcal{Y} }{\delta \Pi^{ij}(x)}-\frac{\delta\mathcal{Y} }{\delta \gamma_{ij}(x)}\frac{\delta\mathcal{X} }{\delta \Pi^{ij}(x)} 
\nonumber\\& \hspace{1.5cm}
+\frac{\delta\mathcal{X} }{\delta \phi^{a}(x)}\frac{\delta\mathcal{Y} }{\delta \Pi_{a}(x)}-\frac{\delta\mathcal{Y} }{\delta \phi^{a}(x)}\frac{\delta\mathcal{X} }{\delta \Pi_{a}(x)} \Bigg] \, .
\end{align}
Thus one can easily examine the following fundamental Poisson brackets which hold between the canonical coordinates and momenta,  
\begin{eqnarray} &&
\lbrace \lambda(\textbf x),  \Pi_{\lambda}(\textbf y)\rbrace=\delta^{(3)}(\textbf x-\textbf y), \nonumber\\ &&
\lbrace \phi^{a}(\textbf x), \Pi_{b}(\textbf y)\rbrace=\delta^{a}_{b}  \delta^{(3)}(\textbf x-\textbf y), \nonumber\\ &&
\lbrace \gamma_{ij}(\textbf x),\Pi^{kl}(\textbf y)\rbrace=\frac{1}{2}(\delta^{k}_{i}\delta^{l}_{j}+\delta^{l}_{i}\delta^{k}_{j})\delta^{(3)}(\textbf x-\textbf y).\label{POb}
\end{eqnarray}
Let us now consider the time evolution of the primary constraint $\Upsilon_{1}\equiv \Pi_{\lambda}\approx 0$ and check its consistency relations. We have
\begin{equation}
\Upsilon_{2}\equiv\partial_{t} \Upsilon_{1}=\{\Upsilon_{1},H_{T}\}=-\frac{G^{ab}}{4\lambda^{2}\sqrt{\gamma}}\Pi_{a}\Pi_{b}+\sqrt{\gamma}(\gamma^{ij} G_{ab}\partial_{i}\phi^{a}\partial_{i}\phi^{b}+1)\approx 0,\label{f11}
\end{equation}
which is exactly the mimetic constraint \eqref{mimeticcons1} written in the ADM decomposition. 

From the above condition one immediately solves for the function $\lambda$,   
\begin{eqnarray} &&
\lambda=(\frac{G^{ab}}{4\sqrt{\gamma}}\Pi_{a}\Pi_{b})^{1/2} \Big[ \sqrt{\gamma}(\gamma^{ij}G_{ab}\partial_{i}\phi^{a}\partial_{j}\phi^{b}+1) \Big]^{-1/2} \, .
\end{eqnarray}

The subsequent consistency relation gives
\begin{eqnarray} &&
\Upsilon_{3}\equiv \partial_{t} \Upsilon_{2}=\{\Upsilon_{2},H_{T}\}=\Upsilon_{3}(\gamma_{ij},\mathcal{N},\beta^{i},\lambda,\Pi_{ij},\Pi_{\mathcal N},\Pi_{i},v_{\lambda})\approx 0,\label{f12}
\end{eqnarray}
which determines the Lagrangian multiplier $v_{\lambda}$ in the terms of phase space variables and then the chain of constraints for primary constraint $\Upsilon_{1}$ terminates here. This means that the set of 2 constraints $\Upsilon_{1}$ and $ \Upsilon_{2} $ 
are second class.
Now, by eliminating $\lambda$ and $\Pi_{\lambda}$ from the constraints $\Upsilon_{1}\approx 0$ and $\Upsilon_{2}\approx 0$,  the dimension of the  phase space reduce and so the reduced total Hamiltonian becomes 
\begin{eqnarray} &&
H_{T}^{R}=\int d^{3}x  \Big[\mathcal{N}(\mathcal{H}_{G}+\mathcal{H}_{M}^{R})+\beta^{i}({\mathcal{H}_{G}}_{i}+{\mathcal{H}_{M}}_{i})+v_{\mathcal{N}}\Pi_{\mathcal{N}}+v^{i}\Pi_{i} \Big]
\end{eqnarray}
where  
\begin{eqnarray}
\mathcal{H}_{M}^{R} \equiv \sqrt{\Pi_{a}\Pi^{a}}\Big(\gamma^{ij}\partial_{i}\phi^{a}\partial_{j}\phi_{a}+1\Big)^{\frac{1}{2}}+\sqrt{\gamma} V(\phi^{a}).
\end{eqnarray}

Now we need to determine the Dirac bracket between the remaining phase space variables  $(\gamma_{ij},\mathcal{N},\beta^{i},\Pi_{ij},\Pi_{\mathcal N},\Pi_{i})$. The Dirac bracket between two phase space functions is defined as
\begin{equation}
\{\mathcal X, \mathcal Y\}_{D} \equiv \{\mathcal X,\mathcal Y\}-\sum_{I,J}\{\mathcal X,\Phi_{I}\}(\mathcal D^{-1})^{IJ}\{\Phi_{J},\mathcal Y\}
\end{equation}
in which $\Phi^{I}$, $I=1,2$ stand for the second class constraints $\Upsilon_{1}\approx0$, $\Upsilon_{2}\approx0$ and $\mathcal D_{IJ}$ is the matrix of the Poisson bracket between these constraints, i.e.,
\begin{eqnarray}
\nonumber \mathcal D_{11}&=&\{{\Upsilon_{1}(\textbf{x})},\Upsilon_{1}(\textbf{y})\}=0\, ,\\
\mathcal D_{12}&=&-\mathcal D_{21}=\{{\Upsilon_{1}(\textbf{x})},\Upsilon_{2}(\textbf{y})\}=\frac{\Pi_{a}(\textbf{x})\Pi^{a}(\textbf{x})}{\sqrt{\gamma}\lambda^{3}(\textbf x)}\delta^{(3)}(\textbf{x}-\textbf{y}) \, ,\\
\nonumber \mathcal D_{22}&=&\{{\Upsilon_{2}(\textbf{x})},\Upsilon_{2}(\textbf{y})\}=\frac{1}{\lambda^{2}(\textbf x)}\Pi_{b}(\textbf{x})\gamma^{ij}(\textbf{y})\partial_{y^{i}}\phi^{b}(\textbf{y})\partial_{y^{j}}\delta^{(3)}(\textbf{x}-\textbf{y}) \, ,\\
\nonumber &-&\frac{1}{\lambda^{2}(\textbf y )}\Pi_{b}(\textbf{y})\gamma^{ij}(\textbf{x})\partial_{x^{i}}\phi^{b}(\textbf{x})\partial_{x^{j}}\delta^{(3)}(\textbf{x}-\textbf{y}) \, .
\end{eqnarray}
Therefore,  the matrix $\mathcal D $ and its inverse $\mathcal D ^{-1}$ can be written schematically as  \cite{chaichian2014mimetic}
\begin{equation}
\mathcal D=\left(\begin{matrix}
0 & A\\
-A & B
\end{matrix}\right), \hspace{1cm} \mathcal D^{-1}=\left(\begin{matrix}
A^{-1}B A^{-1} & -A^{-1}\\
A^{-1} & 0
\end{matrix}\right) \, .
\end{equation}
Keeping the above forms in mind and considering the fact that $\{\gamma_{ij},\Pi_{\lambda}\}=\{\Pi^{ij},\Pi_{\lambda}\}=\{\Pi_{a},\Pi_{\lambda}\}=\{\phi^{a},\Pi_{\lambda}\}=0$, one finds that the Dirac brackets coincides with the Poisson brackets. 

Now we proceed to the analysis of the consistency relation for the primary constraints $\Omega_{1}\equiv \Pi_{\mathcal N}\approx 0$ and $\Gamma_{1}^{i}\equiv \Pi^{i}\approx 0$. The time evolution of these constraints gives the secondary constraints 
\begin{eqnarray} &&
\nonumber \Omega_{2}\equiv\partial_{t} \Omega_{1}=\{\Omega_{1}(\textbf x),H_{T}^{R}(\textbf y)\}=-(\mathcal{H}_{G}+\mathcal{H}_{M}^{R})\delta^{3}(\textbf x-\textbf y) \approx 0,\\&&
\Gamma_{2}^{i}\equiv\partial_{t} \Gamma_{1}^{i}=\{\Gamma_{1}^{i}(\textbf x),H_{T}^{R}(\textbf y)\}=-({\mathcal{H}_{G}}^{i}(\textbf x)+{\mathcal{H}_{M}}^{i}(\textbf x)) \delta^{3}(\textbf x-\textbf y)\approx 0 \, .
\end{eqnarray}
Requiring these constraints to be time independent yields 
\begin{eqnarray} &&
\nonumber \Omega_{3}=\partial_{t}\Omega_{2}=\{\Omega_{2}(\textbf x),{H}_{T}^{R}(\textbf y)\}=-\mathcal{N}\{\Omega_{2}(\textbf x),\Omega_{2}(\textbf y)\}-\beta_{i}\{\Omega_{2}(\textbf x),\Gamma_{2}^{i}(\textbf y) \}\approx 0,\\&&
\Gamma_{3}^{i}=\partial_{t} \Gamma_{2}^{i}= \{\Gamma_{2}^{i}(\textbf{x}),H_{T}^{R}(\textbf y) \}=-\mathcal{N}\{\Gamma_{2}^{i}(\textbf x),\Omega_{2}(\textbf y)\}-\beta_{j}\{\Gamma_{2}^{i}(\textbf x),\Gamma_{2}^{j}(\textbf y) \}\approx 0.
\end{eqnarray}
where
\begin{eqnarray} &&
\nonumber \{ \Omega_{2}(\textbf x), \Omega_{2}(\textbf y)\}= \Gamma_{2}^{i}(\textbf y)\partial_{x^{i}}\delta^{(3)}(\textbf x-\textbf y)- \Gamma_{2}^{j}(\textbf  x)\partial_{y^{j}}\delta^{(3)}(\textbf x-\textbf y)\approx 0,\\&&
\{\Omega_{2}(\textbf x),\Gamma_{2}^{i}(\textbf y) \}=-\Omega_{2} \partial_{x_{i}}\delta^{(3)}(\textbf x-\textbf y)\approx 0,\\&&
\nonumber \{\Gamma_{2}^{i}(\textbf x),\Gamma_{2}^{j}(\textbf y)\}=\Gamma_{2}^{i}(\textbf y)\partial_{{x}_{j}}\delta^{(3)}(\textbf x-\textbf y)-\Gamma_{2}^{j}(\textbf x)\partial_{y_{i}}\delta^{(3)}(\textbf x-\textbf y)\approx 0.
\end{eqnarray}
Clearly, the above Poisson algebras vanish on the constraint surface (see Append. \ref{AppA} for more details.). Indeed, the time evolution of the secondary constraints $\Omega_{2}$ and $\Gamma_{2}^{i}$ determine none of the Lagrangian multipliers and do not generate any additional constraints. Consequently   
these eight constraints  $ \Omega_{1} $, $ \Omega_{2} $, $\Gamma_{1}^{i} $, and $ \Gamma_{2}^{i} $ are all first class constraints, which are interpreted as the generators of diffeomorphism.

Now the DOFs in phase space can be read off from the master formula of the constrained system as \cite{dirac1964lectures} 
\begin{eqnarray} &&
\text{DOF} = N-2  \#  \text{1st Class} -\#  \text{2nd Class},\label{mf}
\end{eqnarray}
in which $ N $ is the total number of phase space variables. In our model, we have
twenty phase space variables containing $(\mathcal N, \beta^{i},\gamma_{ij}, \Pi_{\mathcal N}, \Pi^{i},\Pi_{ij})$, two variables for $( \lambda , \Pi_{\lambda}) $, and $ \mathcal M $ total number of the conjugate pair  $ (\phi^{a}, \Pi_{a}) $. Correspondingly, 
the number of DOFs is obtained to be
\begin{eqnarray}\label{DOF}
\text{DOF}=(20+2+\mathcal M)-16-2=4+\mathcal M,
\end{eqnarray}
which corresponds to  $(4+\mathcal M)/2$ physical degrees of freedom in the configuration space. 

Note that $\mathcal{M}/2$ represents the   dimension of the field space manifold or equivalently the number of  scalar fields. Thus, in addition to the two gravitational degrees of freedom of general relativity, there exists $\mathcal{M}/2$ extra physical degree of freedom. This finding implies that the theory is  free from the so-called Ostrogradsky ghost \cite{Woodard:2015zca}. In fact, the Ostrogradsky ghost arises when the higher derivative terms increase the number of degrees of freedom for the system under consideration. 

As an example, for two-field mimetic case spanning a 2D curved field space ($\mathcal{M}=2$), we have 4 DOFs corresponding to one adiabatic mode, one entropy mode and two tensor modes as verified  perturbatively in previous section. 

\section{Conclusions}
\label{sec5}

In this paper we have extended  the idea of mimetic gravity to multi-field setup with the curved field space manifold. Motivated from the single field case, we have constructed the multi-field mimetic setup via the conformal transformation between  the physical 
and the original auxiliary metrics. Solving the eigentensor equation for the Jacobian of such a transformation, we have found the associated eigentensors and eigenvalues. The multi-field generalization of the mimetic scenario can be interpreted as the singular limit of the conformal transformation by demanding that the kinetic type eigenvalues to be zero.

At the cosmological background, as expected, the energy density of the 
multi-field mimetic scenario indeed plays the role of dark mater similar to the original single field setup. At the perturbation level, we have employed the kinematic basis in which the perturbations are decomposed into the tangential and the perpendicular to the field space trajectory.  By considering the perturbations in the kinematic basis, we performed explicit perturbation analysis
in the spatially flat gauge for two-field mimetic case. We have found that the perturbation mode perpendicular to background trajectory in the field space manifold, $u_{N}$, is excited, whereas the perturbation mode tangential to the background trajectory, $u_{T}$, does not propagate. More precisely, the entropy mode, originated from the extra scalar field in
our model, propagates with the sound speed equal to unity, whereas the sound speed for the adiabatic mode is zero. This is consistent with  the fact that the mimetic background describes a dark-matter fluid. In addition, whether or not the entropy perturbation is healthy directly depends on the signature of the field-space metric $G_{ab}$. In the case of the field space with an Euclidean signature, the entropy mode is healthy, whereas in the case of the Lorentzian manifold, the entropy perturbation suffers from the ghost and gradient instabilities.


Although one of the modes does not propagate at the linear perturbation level but it contributes to the physical DOFs of the model. We further confirmed this result by performing the full non-linear Hamiltonian analysis of the multi-field mimetic theory and verified the correct number of DOFs necessary to prevent the presence of the Ostrogradsky-type ghost.

\subsection*{Acknowledgments}

We would like to thank  Mohammad Ali Gorji for insightful discussions. H. F. and A. T. acknowledge the partial support from the ``Saramadan'' federation of Iran.


\vspace{0.5cm}

\appendix

\section{Constraint algebra}\label{AppA}
Using the Poisson bracket relations \eqref{poissonb}, here we check the constraint $\Omega_{2}\equiv -\mathcal{H}_{G}-\mathcal{H}_{M}^{R}$ and $\Gamma_{2}^{i}\equiv -\mathcal{H}_{G}^{i}-\mathcal{H}_{M}^{i}$ in a closed algebra \cite{dirac1964lectures,bojowald2010canonical}. The gravity part $\mathcal{H}_{G}^{i}$  are the generators of 3-dimensional diffeomorphism and the other gravity part $\mathcal{H}_{G}$ is a scalar with respect to spatial diffeomorphism \cite{Gong:2016qpq}, then $\mathcal{H}_{G}$ and $\mathcal{H}_{G}^{i}$ satisfy 
\begin{eqnarray} && \nonumber 
\{\mathcal{H}_{G}(\textbf x),\mathcal{H}_{G}(\textbf y)\}=\mathcal{H}_{G}^{i}(\textbf y)\partial_{x^{i}}\delta^{(3)}(\textbf x-\textbf y)-\mathcal{H}_{G}^{j}(\textbf x)\partial_{y^{j}}\delta^{(3)}(\textbf x-\textbf y),\\&& 
\{\mathcal{H}_{G}(\textbf x),\mathcal{H}_{G}^{i}(\textbf y)\}=-\mathcal{H}_{G}\partial_{i}\delta^{(3)}(\textbf x-\textbf y),\\&& \nonumber 
\{\mathcal{H}_{G}^{i}(\textbf x),\mathcal{H}_{G}^{j}(\textbf y)\}=\mathcal{H}_{G}^{i}(\textbf y)\partial_{{x}_{j}}\delta^{(3)}(\textbf x-\textbf y)-\mathcal{H}_{G}^{j}(\textbf x)\partial_{{y}_{i}}\delta^{(3)}(\textbf x-\textbf y).
\end{eqnarray}
For the matter part, i.e. $\mathcal{H}^{R}_{M}$ and $\mathcal{H}^{i}_{M}$, similarly one obtains 
\begin{eqnarray} &&\nonumber 
\{\mathcal{H}^{R}_{M}(\textbf x),\mathcal{H}^{R}_{M}(\textbf y)\}=\mathcal H^{ i}_{M}(\textbf y)\partial_{x^{i}}\delta^{(3)}(\textbf x-\textbf y)-\mathcal{H}_{M}^{j}(\textbf x)\partial_{y^{j}}\delta^{(3)}(\textbf x-\textbf y),\\&&
\{\mathcal{H}^{R}_{M}(\textbf x),\mathcal{H}_{M}^{i}(\textbf y) \}=-\mathcal{H}^{R}_{M}\partial_{x_{i}}\delta^{(3)}(\textbf x-\textbf y),\\&& \nonumber 
\{\mathcal{H}_{M}^{i}(\textbf x),\mathcal{H}_{M}^{j}(\textbf y)\}=\mathcal{H}_{M}^{i}(\textbf y)\partial_{x_{j}}\delta^{(3)}(\textbf x-\textbf y)-\mathcal{H}_{M}^{j}(\textbf x)\partial_{{y}_{i}}\delta^{(3)}(\textbf x-\textbf y).
\end{eqnarray}
Therefore, the total constraints $\Omega_{2}\equiv -\mathcal{H}_{G}-\mathcal{H}_{M}^{R}$ and $\Gamma_{2}^{i}\equiv -\mathcal{H}_{G}^{i}-\mathcal{H}_{M}^{i}$ also form the closed algebra,
 \begin{eqnarray} &&
 \nonumber \{ \Omega_{2}(\textbf x), \Omega_{2}(\textbf y)\}= \Gamma_{2}^{i}(\textbf y)\partial_{x^{i}}\delta^{(3)}(\textbf x-\textbf y)- \Gamma_{2}^{j}(\textbf  x)\partial_{y^{j}}\delta^{(3)}(\textbf x-\textbf y),\\&&
 \{\Omega_{2}(\textbf x),\Gamma_{2}^{i}(\textbf y) \}=-\Omega_{2} \partial_{x_{i}}\delta^{(3)}(\textbf x-\textbf y),\\&&
 \nonumber \{\Gamma_{2}^{i}(\textbf x),\Gamma_{2}^{j}(\textbf y)\}=\Gamma_{2}^{i}(\textbf y)\partial_{{x}_{j}}\delta^{(3)}(\textbf x-\textbf y)-\Gamma_{2}^{j}(\textbf x)\partial_{y_{i}}\delta^{(3)}(\textbf x-\textbf y).
 \end{eqnarray}

\section{Quadratic action in the comoving gauge}\label{AppB}

In comoving gauge  $\psi$ is turned on in the scalar perturbations \eqref{pertubation}. In this respect, the equation of motion for the non-dynamical mode $B$ leads to the following constraint,
\begin{equation}
	\alpha=\frac{\dot{\psi}}{H}+\epsilon H Q_{a}\dot{\phi}_{0}^{a} \, .
\end{equation}
By imposing the above relation in the corresponding quadratic action, one obtains
\begin{eqnarray}\label{comovingac}
\nonumber S^{2} &=& \int d^3 x dt a^3 \Big[M_P^2 \epsilon H^2 G_{bc}D_{t}Q^{b} D_{t}Q^{c} -2 M_P^2  \epsilon^2 H^3 Q_{b} D_{t} Q_{c} \dot \phi_{0}^{b} \dot \phi_{0}^{c}-M^{2}_{bc} Q^{b} Q^{c}\\
\nonumber & -& \frac{M_P^2}{a^2} \epsilon H^2 G_{bc} \partial_{i}  Q^{b}\partial^{i} Q^{c}+ 2 \delta \lambda \Big(\epsilon H Q_{b} \dot \phi_{0}^{b}-D_{t} Q_{b} \dot \phi_{0}^{b}+\frac{\dot{\psi}}{H}\Big)+M_P^2 \epsilon \dot{\psi}^2\\
\nonumber &-&\psi \Big(3 Q^{b}V_{b}-6 M_P^2 \epsilon H^2 D_{t}Q_{b}\dot{\phi}_{0}^{b}\Big)-\frac{M_P^2}{a^2} \epsilon \partial^2 \psi\Big( \psi+2 H Q_{b} \dot{\phi}_{0}^{b}+2 \frac{\dot{\psi}}{H}\Big) \\
 &-&\dot{\psi}\Big(\frac{V_{d}Q^{d}}{H}-2 M_P^2 \epsilon^2 H^2 Q_{b}\dot{\phi_{0}^{b}}+2 M_P^2 \epsilon H D_{t}Q_{b} \dot{\phi}_{0}^{b}\Big)\Big] \, ,
\end{eqnarray}
where the mass matrix $M_{ab}$ was defined in Eq. \eqref{massmatrix}. Now, we can decompose the variable $ Q^a$ into the directions along and orthogonal to time evolution~\cite{Achucarro:2012sm} as 
\begin{equation}
\label{eq:decomposition2}
 Q^a = Q^a_\bot + \dot\phi^a_0\tilde \pi \, ,
\end{equation}
with the orthogonality condition $G_{ab} \dot{\phi}_{0}^{a}Q^a_\bot=0$.  In the comoving gauge we impose $\tilde \pi=0$. It should be noted that the $\tilde \pi$ mode is the fluctuation in the direction of the time translation, and is interpreted as the Goldstone mode appearing from the spontaneous breaking of time translation invariant \cite{Cheung:2007st}. Moreover, the orthogonal modes, $Q^a_\bot$, are gauge invariant quantities and are usually called ``isocurvature'' modes \cite{Gordon:2000hv}. 
Similar to the single field inflation, one can introduce the Mukhanov-Sasaki variable as \cite{Gong:2016qmq} 
\begin{equation}
	\tilde{Q}^{a}\equiv Q^{a}-\frac{\dot{\phi}_{0}^a}{H} \psi=Q^a_\bot - \frac{\dot\phi^a_0}{H}(\psi- H\tilde \pi )  \equiv Q^a_\bot - \frac{\dot\phi^a_0}{H}  \pi
\end{equation} 
or equivalently 
\begin{equation}\label{quan}
Q^{a}\equiv  Q^a_\bot-\frac{\dot\phi^a_0}{H} ( \pi- \psi) \, .
\end{equation}
In two-field case, due to the orthogonality condition, the mode $Q_\bot^a$ is proportional to the normal vector $N^a$, i.e.,  $Q_\bot^a \propto N^a$, and one takes the amplitude of $Q_\bot$ as the isocurvature field $\mathcal{F}$ \cite{Gong:2016qmq}. Moreover, in comoving gauge ($\tilde{\pi}=0$) in which  $\psi=\mathcal R$, one can replace $\pi$ simply with the curvature perturbation $\mathcal R$.

In this regard, the variation of the quadratic action \eqref{comovingac} with respect to $\delta \lambda$, in comoving gauge, yields the constraint
\begin{equation}
\label{Rdot-F}
\dot{\mathcal R}=- H \dot \theta \mathcal{F} \,.
\end{equation}
By imposing the above constraint and Eq. \eqref{quan} into the quadratic action \eqref{comovingac}, finally we arrive  at the quadratic action in comoving gauge,
\begin{eqnarray}
\label{action-comoving}
\nonumber S^{2}&=&\int {\rm d} x^3 \   {\rm d}t \  a^3 M_P^2 \epsilon H^2 \Big[sgn(\pm 1)\Big(\dot{\mathcal{F}}^2-\frac{1}{a^2} (\partial \mathcal{F})^2\Big)-(\frac{M_{NN}^2}{M_P^2 \epsilon H^2} +2 \dot{\theta}^2)\mathcal{F}^2\\
&+& \frac{2\dot{\theta}}{a^2 \epsilon H^2}  \mathcal{F} \partial^2 \mathcal{R}+\frac{1}{\epsilon H^2 a^2} (\partial \mathcal{R})^2\Big] \, .
\end{eqnarray}
It implies that the curvature perturbation $\mathcal{R}$ does not propagate in our setup. 
Clearly, the stability of perturbations depends on the signature of the metric $G_{ab}$. In the case of the field space with an Euclidean signature ($sgn(+1)=1$), the isocurvature mode does not suffer from ghost and gradient instabilities, whereas in the case of the Lorentzian manifold with $sgn(-1)=-1$, the isocurvature perturbation is pathological. In 2D flat field space, our results are in agreement with those in \cite{Shen:2019nyp}.

\vspace{1cm}
\bibliographystyle{JHEP}
\bibliography{references}

\end{document}